\documentclass[amsmath,superscriptaddress,amssymb,aps,onecolumn,pra]{revtex4}
\usepackage{graphicx}
\usepackage{soul}
\usepackage[colorlinks=true,citecolor=blue,linkcolor=magenta]{hyperref}
\usepackage[usenames]{color}

\def \be{\begin{equation}}
\def \ee{\end{equation}}
\def \ba{\begin{array}}
\def \ea{\end{array}}
\def \bea{\begin{eqnarray}}
\def \eea{\end{eqnarray}}

\newcommand{\ket}[1]{\left\vert #1 \right\rangle}
\newcommand{\bra}[1]{\left\langle #1 \right\vert}
\newcommand{\ketbra}[2]{\ket{ #1}\bra{ #2} }
\newcommand{\bla}[1]{\left( #1 \right)}
\newcommand{\blb}[1]{\left[ #1 \right]}

\begin{document}

\title{Robust dynamical decoupling with concatenated continuous driving}

\author{Jianming Cai}
\affiliation{Institut f\"{u}r Theoretische Physik, Albert-Einstein Allee 11, Universit\"{a}t Ulm, 89069 Ulm, Germany}
\affiliation{Center for Integrated Quantum Science and Technology, Universit\"{a}t Ulm, 89069 Ulm, Germany}
\author{ Boris Naydenov}
\affiliation{Center for Integrated Quantum Science and Technology, Universit\"{a}t Ulm, 89069 Ulm, Germany}
\affiliation{Institut f\"{u}r Quantenoptik, Albert-Einstein Allee 11, Universit\"{a}t Ulm, 89069 Ulm, Germany}
\author{Rainer Pfeiffer}
\affiliation{Center for Integrated Quantum Science and Technology, Universit\"{a}t Ulm, 89069 Ulm, Germany}
\affiliation{Institut f\"{u}r Quantenoptik, Albert-Einstein Allee 11, Universit\"{a}t Ulm, 89069 Ulm, Germany}
\author{Liam P. McGuinness}
\affiliation{Center for Integrated Quantum Science and Technology, Universit\"{a}t Ulm, 89069 Ulm, Germany}
\affiliation{Institut f\"{u}r Quantenoptik, Albert-Einstein Allee 11, Universit\"{a}t Ulm, 89069 Ulm, Germany}
\author{Kay D. Jahnke}
\affiliation{Center for Integrated Quantum Science and Technology, Universit\"{a}t Ulm, 89069 Ulm, Germany}
\affiliation{Institut f\"{u}r Quantenoptik, Albert-Einstein Allee 11, Universit\"{a}t Ulm, 89069 Ulm, Germany}
\author{Fedor Jelezko}
\affiliation{Center for Integrated Quantum Science and Technology, Universit\"{a}t Ulm, 89069 Ulm, Germany}
\affiliation{Institut f\"{u}r Quantenoptik, Albert-Einstein Allee 11, Universit\"{a}t Ulm, 89069 Ulm, Germany}
\author{Martin B. Plenio}
\affiliation{Center for Integrated Quantum Science and Technology, Universit\"{a}t Ulm, 89069 Ulm, Germany}
\affiliation{Institut f\"{u}r Theoretische Physik, Albert-Einstein Allee 11, Universit\"{a}t Ulm, 89069 Ulm, Germany}
\author{Alex Retzker}
\affiliation{Institut f\"{u}r Theoretische Physik, Albert-Einstein Allee 11, Universit\"{a}t Ulm, 89069 Ulm, Germany}
\affiliation{Racah Institute of Physics, The Hebrew University of Jerusalem, Jerusalem 91904, Israel}

\date{\today}

\begin{abstract}
The loss of coherence is one of the main obstacles for the implementation of quantum information
processing. The efficiency of dynamical decoupling schemes, which have been introduced to address
this problem, is limited itself by the fluctuations in the driving fields which will themselves introduce
noise. We address this challenge by introducing the concept of concatenated continuous dynamical
decoupling, which can overcome not only external magnetic noise but also noise due to fluctuations in driving fields. We show theoretically that this approach can achieve relaxation limited  coherence times, and demonstrate experimentally that already the most basic implementation of this concept yields an order of magnitude improvement of the decoherence time for the electron spin of nitrogen vacancy centers in diamond. The proposed scheme can be  applied to a wide variety of other physical systems including,  trapped atoms and ions, quantum dots, and may be combined with other quantum technologies challenges such as quantum sensing and quantum information processing.
\end{abstract}

\maketitle

\section{Introduction}

Coherent control of quantum systems has opened a promising route towards novel quantum devices for quantum technologies, such as quantum information processing, quantum metrology and quantum sensing \cite{Maze08,Bal08,Jones09,Hanson11}. The performance of such quantum devices critically depends on coherence times of their constituent quantum systems which, in turn, are limited by uncontrolled interactions with their surrounding environment. This results in a challenging but fundamentally important task in current quantum experiments, namely how to protect individual quantum states from decoherence by their environment while retaining the ability to control the quantum dynamics of the system, in particular in solid state systems with characteristic complex environments. For slow environmental noise a successful strategy is dynamical decoupling due to rapid sequences of intense pulses of electromagnetic radiation which has been applied successfully in NMR\cite{Hahn50}. Significant progress has been made with the theoretical proposals of various dynamical decoupling \cite{Viola98,Pas04,Lidar05,Uhrig07,Rabl,Ber11,Alex11,Fan07}, and their experimental demonstrations \cite{Bier09,Du09,Lange10,Cory10,Timoney11,Suter11,Vitali09,Nay11,Sar12,Lar11}.

The recently developed dynamical decoupling schemes that require only continuous oscillatory driving fields \cite{Pas04,Fan07}, inherit the advantages of standard dynamical decoupling, namely requiring no encoding overhead, no quantum measurements, and no feedback controls. Moreover, they are easier to realize experimentally and are more naturally combined with other quantum information tasks, such as the implementation of high fidelity quantum gates \cite{Rabl,Timoney11,Ber11,Alex11}. In principle, one can apply continuous driving to reduce the noise suffered by a qubit considerably simply by increasing their intensity. However, random and systematic fluctuations which are inevitably present in the driving field itself will ultimately limit the efficiency of dynamical decoupling. The deleterious effect of driving field fluctuations in particular will become significant when employing strong driving fields to achieve ultralong coherence times. This is important in all applications in which environmental noise cannot be avoided (e.g. sensing in biological environments) and thus represents a fundamental obstacle. Overcoming the limitations imposed by driving field fluctuations and thus extending coherence times further represents a key step towards the construction of quantum memory \cite{Fuchs11}, highly sensitive nano-scale magnetometers \cite{Maze08,Bal08,Hanson11} and error-resistant quantum operation \cite{Alex11}. It will also be of particular interest for $\mbox{T}_1$ limited Rabi-type magnetic resonance imaging, the resolution of which highly depends on the stability of microwave driving fields \cite{Fedder10,DS11}.

In this work, we address this challenge by introducing the scheme of concatenated continuous decoupling (CCD), which can significantly extend coherence times by protecting against driving field fluctuations. As a proof-of-principle, we implement a second-order CCD with a single NV center in diamond where a second, weaker driving field, reduces the impact of the amplitude fluctuations of the first-order driving field. We demonstrate experimentally that CCD schemes with only a weak second driving field can already increase coherence times by an order of magnitude as compared to standard schemes based on a single drive.

\section{Concatenated continuous dynamical decoupling}

\begin{figure}[t]
\begin{center}
\includegraphics[width=9cm]{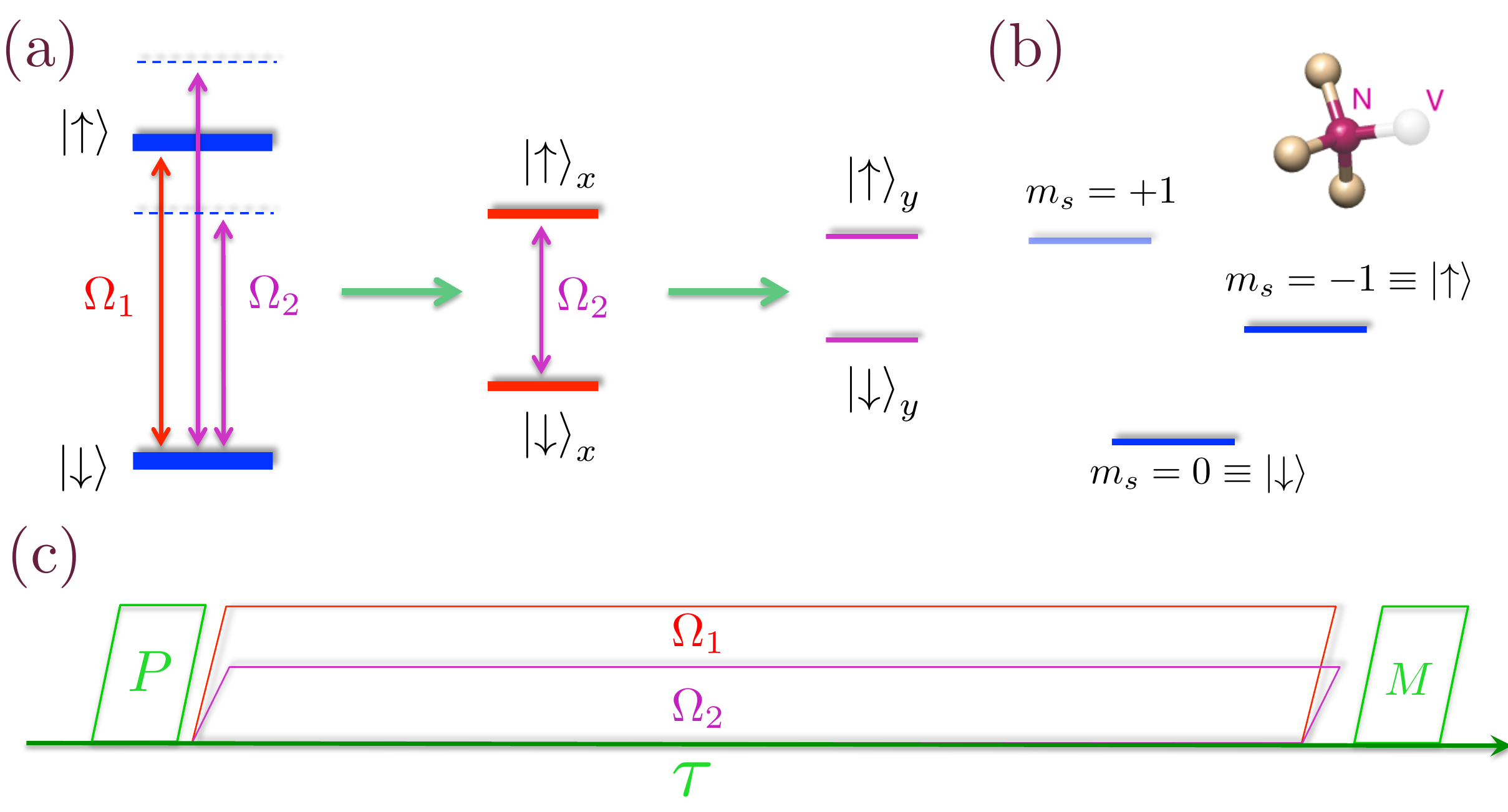}
\end{center}
\caption{(Color online) \textbf{a}: Second-order concatenated  continuous dynamical decoupling scheme: The first-order driving field with the frequency $\omega$ and the Rabi frequency $\Omega_1$ creates the first-order dressed states $\ket{\uparrow}_x$ and $\ket{\downarrow}_x$, which suffer less from the  dephasing effect of environmental noise. These dressed states are however subject to the fluctuation in the amplitude $\Omega_1$ of the driving field. The additional second-order driving field with the amplitude $\Omega_2$ (which is generally smaller than $\Omega_1$) has a detuning $\pm \Omega_1$ and a relative phase $\frac{\pi}{2}$ with respect to the first-order driving field. This second order driving field leads to second-order dressed states $\ket{\uparrow}$
and $\ket{\downarrow}$ which are protected against the intensity fluctuations of the first-order drive and are thus suffering much reduced energy fluctuations. This scheme may be iterated further to n-th order. \textbf{b}:  The diagram for the energy levels of the NV center electron spin. The NV spin triplet electronic ground state is splitted by an applied magnetic field. The effective two-level system used in our experiment is formed by the spin sublevels $m_s = 0$ (labeled as $\ket{\downarrow}$) and $m_s = -1$ (labeled as $\ket{\uparrow}$). \textbf{c}: The procedure for the Ramsey experiment with the second-order dressed qubit: the NV electron spin is first polarized into the $m_s =0$ sublevel, which is a coherent superposition of the second-order dressed qubit $\ket{\uparrow}_y$ and $\ket{\downarrow}_y$;  the first- and second-order driving fields are simultaneously switched on for time $\tau$ followed by the optically readout of the population of $m_s=0$ via spin-dependent fluorescence.}
\label{setup}
\end{figure}

We start by considering a two-level quantum system with eigenstates $\ket{\uparrow}, \ket{\downarrow}$, see Fig.\ref{setup}(a), and Hamiltonian $H_0=\frac{\hbar\omega}{2} \bla{\ketbra{\uparrow}{\uparrow}-\ketbra{\downarrow}{\downarrow}}$. Environmental noise causes fluctuations to the energies $\omega_{\uparrow}$, $\omega_{\downarrow}$ and thus the loss of coherence. To counter such effects, we can apply a driving field on resonance with the energy gap $\hbar \omega$ between $\ket{\uparrow}$ and $\ket{\downarrow}$ as
\begin{equation}
H_{d_1}= \hbar\Omega_1 \cos{(\omega t)} \sigma_x
\end{equation}
where $ \sigma_x=\ketbra{\uparrow}{\downarrow}+\ketbra{\downarrow}{\uparrow}$. In the interaction picture with respect to $H_0$ and with rotating wave approximation, we find $H_I^{(1)}=\frac{\hbar\Omega_1}{2} \sigma_x$, and its eigenstates $\ket{\uparrow}_x=\frac{1}{\sqrt{2}}\bla{\ket{\uparrow} +\ket{\downarrow}}$ and $\ket{\downarrow}_x=\frac{1}{\sqrt{2}}\bla{\ket{\uparrow} -\ket{\downarrow}}$ are the dressed states \cite{Gry92}. In this basis, the effect of dephasing noise now induces transitions among these dressed states, which are suppressed by an energy penalty as long as the noise power spectrum at the resonance frequency is negligible \cite{Timoney11}. The decoupling efficiency will be limited if the noise has a wide range of frequencies, while it can be very efficient for slow baths e.g. in diamond \cite{Hanson08,Rein12}. The above analysis is based on the assumption that the amplitude of the driving fields is stable. In realistic experiments however, the intensity of the driving fields will fluctuate owing to limited stability of microwave sources and amplifiers (the frequency instead can be relatively much more stable), and thus cause fluctuations of the energies of the dressed states. Achievable coherence times of the dressed qubit using a single drive are thus ultimately limited by the stability of the driving fields \cite{Fedder10}, which appears as the fast decay of Rabi oscillation.

The principal idea of CCD is to provide a concatenated set of continuous driving fields with decreasing intensities (and thus smaller absolute value of fluctuation) such that each new driving field protects against the fluctuations of the driving field at the preceding level. For example, fluctuations in the amplitude of the first-order driving can be suppressed by applying a second-order driving field as follows
\begin{equation}
%H_{d_2}=\hbar\Omega_2  \cos{\blb{\bla{\omega+\Omega_1}t+\varphi}}\sigma_x.
H_{d_2}=2\hbar \Omega_2 \cos(\omega t+\frac{\pi}{2}) \cos(\Omega_1 t) \sigma_x.
\label{Hd2-m}
\end{equation}
This second-order driving field is on resonance with the the energy gap of the first-order dressed state, see Fig.\ref{setup}(a), which actually describes rotation about the axis $\hat{y}$ in the interaction picture (see Appendix for details) and thus plays the role of decoupling the first-order dressed states from the fluctuation of $\Omega_1$. With these two driving fields as $H_{d_1}$ and $H_{d_2}$, we find the effective Hamiltonian in the second-order interaction picture (see Appendix for details) as
\begin{equation}
H_I^{(2)}=\frac{\hbar\Omega_2}{2}  \sigma_y
\end{equation}
where $\sigma_y=-i \ketbra{\uparrow}{\downarrow}+i \ketbra{\downarrow}{\uparrow}$ and the second-order dressed states are
$\ket{\uparrow}_y=\sqrt{\frac{1}{2}} \bla{\ket{\uparrow}+i\ket{\downarrow}}$ and  $\ket{\downarrow}_y=\sqrt{\frac{1}{2}} \bla{\ket{\uparrow}-i\ket{\downarrow}}$, which can be used to encode and store quantum information.

In general, we can apply n-th order continuous driving fields on the condition that two subsequent drivings describe rotations about orthogonal axes ($\hat{x}$ and $\hat{y}$) in the corresponding interaction picture. The higher-order driving fields can be explicitly written as
%For example, the third- and fourth-order driving fields are
%\begin{eqnarray}
%H_{d_3} &=& 2\hbar\Omega_3 \cos(\omega t) \cos(\Omega_2 t) \sigma_x,\\
%H_{d_4} &= &4\hbar\Omega_4 \cos(\Omega_1 t )\cos(\Omega_3 t) \cos(\omega t+\frac{\pi}{2})\sigma_x.
%\end{eqnarray}
\begin{eqnarray}
H_{d_{2k+1}} &=& 2^{k}\hbar\Omega_{2k+1}   \prod_{j=1}^{k}\cos(\Omega_{2j} t ) \cos(\omega t) \sigma_x,\label{Hd2k1}\\
H_{d_{2k}} &= &2 ^{k}\hbar\Omega_{2k} \prod_{j=1}^{k}\cos(\Omega_{2j-1} t ) \cos(\omega t+\frac{\pi}{2})\sigma_x. \label{Hd2k}
\end{eqnarray}
We would like to stress that the achievable coherence time is limited by the $\mbox{T}_1$ time which can usually be achieved using a fourth order scheme in the context of NV centers in diamond.
%We remark that usually it may already be enough to significantly extend the coherence time using up to the fourth-order driving fields.
%We remark that the relative phase does not need to take a specific value, once it is fixed the corresponding second-order dressed %state can serve as a robust encoded qubit see SI for details.
The decoherence of the n-th order dressed qubit stems dominantly from the fluctuation of the n-th order driving field. As long as the orthogonality of two consecutive driving field is satisfied, the (n+1)-th driving field protects against the noise of the n-th driving field. Hence with a concatenated scheme in which subsequent driving fields have decreasing intensities, the effective dephasing will be sequentially suppressed and coherence times can be extended significantly.

\begin{figure}[t]
\begin{center}
\begin{minipage}{12cm}
\hspace{-0.8cm}
\includegraphics[width=6cm]{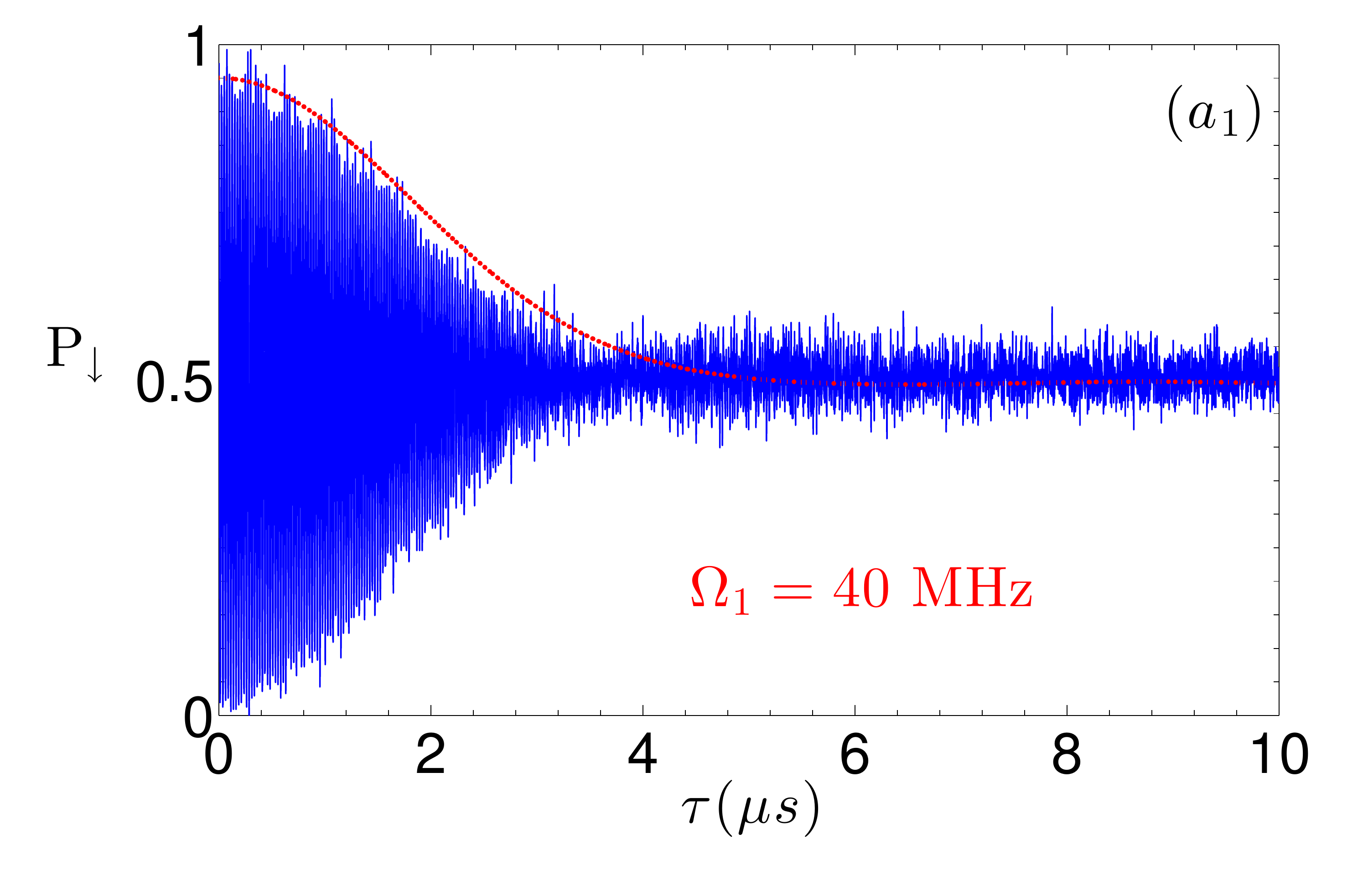}
%\end{minipage}
%\begin{minipage}{9cm}
\hspace{-0.4cm}
\includegraphics[width=6cm]{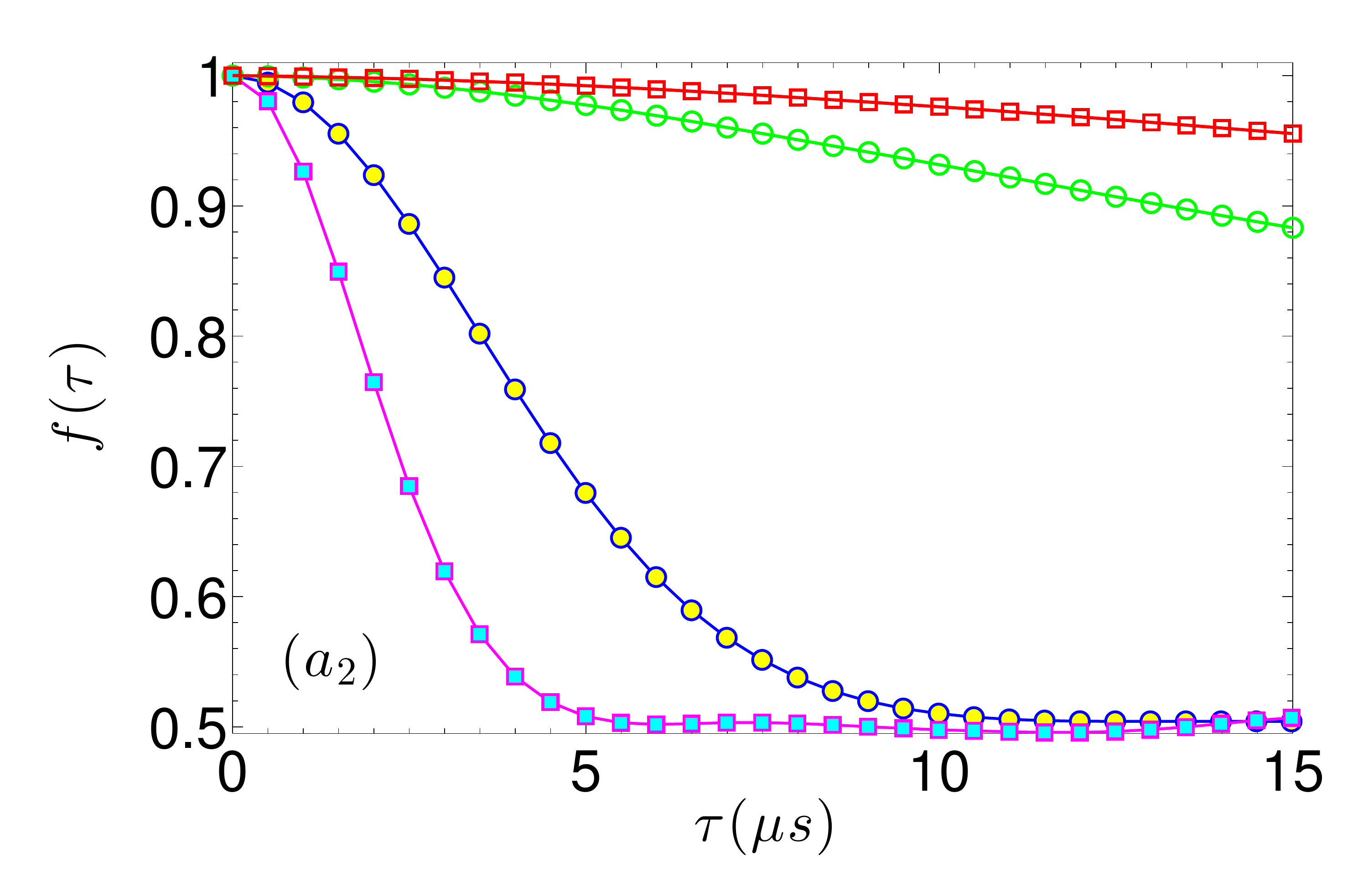}
\end{minipage}
\begin{minipage}{12cm}
\hspace{-0.6cm}
\includegraphics[width=12cm]{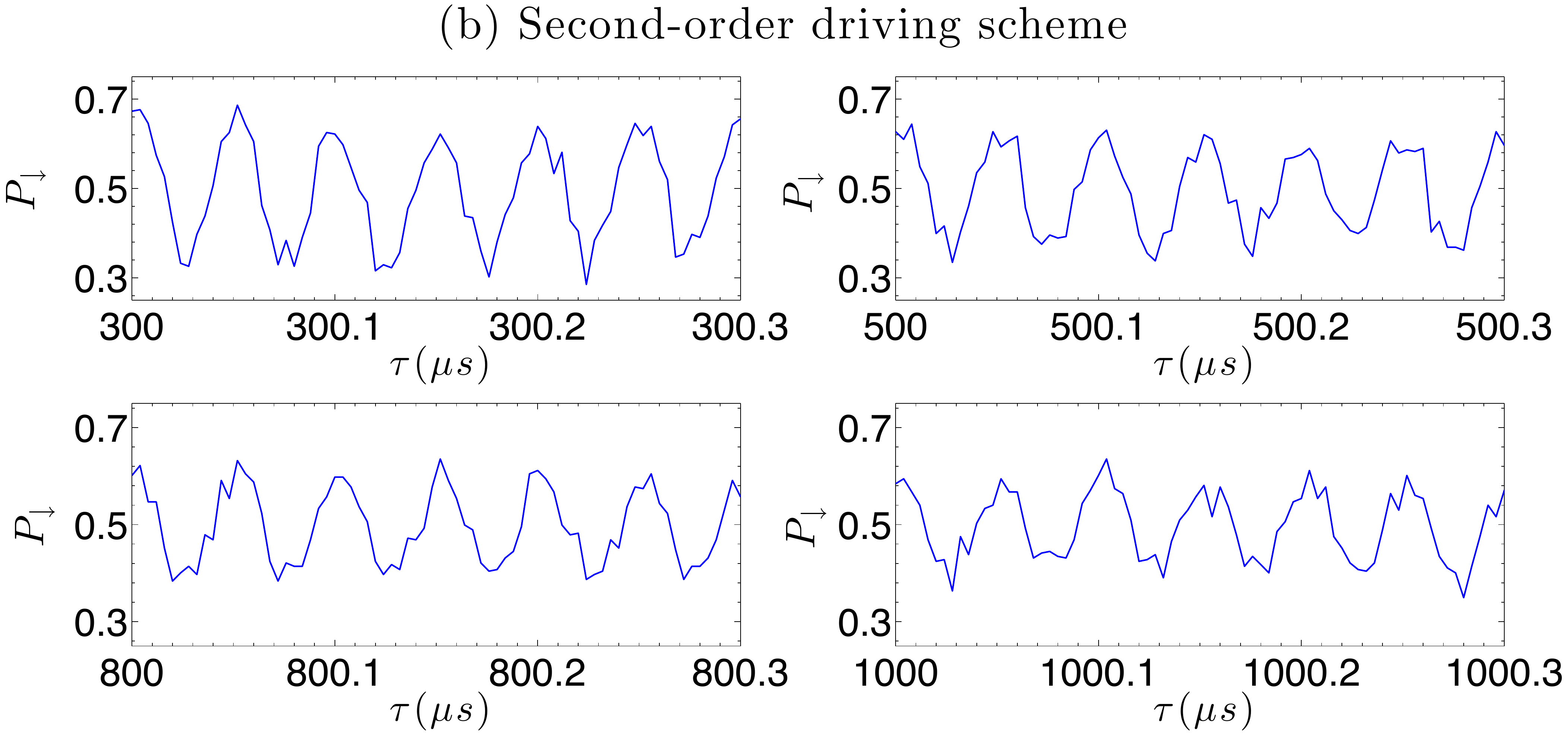}
\end{minipage}
\end{center}
\caption{(Color online) $\mathbf{a_1}$: Coherent driven oscillation of NV center with a single microwave field. The Rabi frequency is $\Omega_1=40 \mbox{MHz}$. The blue curve is the experimental data, and the red one is the decay envelope from numerical simulation. We thereby estimate that the relative amplitude fluctuation is about $2.4\times 10^{-3}$ (i.e. $98 \mbox{kHz}$), and the decay time is about $2.3 \mu s$. $\mathbf{a_2}$: Numerical simulation for the decay of Rabi oscillation under different drivings: $40 \mbox{MHz}$ (purple, rectangle) and $20 \mbox{MHz}$ (blue, circle). For comparison, we also plot the results assuming that there is no microwave fluctuation:  $40 \mbox{MHz}$ (red, rectangle) and $20 \mbox{MHz}$ (green circle). \textbf{b}: Persistent Rabi oscillation by adding a second-order driving field, the intensity of which is 10 times weaker than the first driving ($\Omega_1=20 \mbox{MHz}$). The four panels shows the very slow decay of Rabi oscillation beyond $1000 \mu s$.
}\label{Rabi}
\end{figure}

\section{Experimental demonstration of CCD Scheme}

We have used NV centers in diamond to demonstrate the working principle and efficiency of our concatenated continuous dynamical decoupling scheme. As a promising candidate physical system for modern quantum technologies, NV centers have been used to demonstrate basic quantum information processing protocols \cite{Childress06}, as well as ultrasensitive magnetometry and nano-scale imaging at room temperature \cite{Jelezko04,Jiang09,Neumann10}. The electronic ground state of NV center is a spin triplet with three sublevels with magnetic quantum numbers $m_s=0$ and $m_s=\pm 1$, and the zero field splitting is $\sim 2.87 \mbox{GHz}$ \cite{Wra06}, see Fig.\ref{setup} (b). We apply an additional magnetic field along the axis of the NV center to split the energy levels of $m_s=\pm 1$. The two electronic transition frequencies corresponding to $m_s=0 \rightarrow m_s=\pm 1$ are determined to be $2042 \mbox{MHz}$ and $3696 \mbox{MHz}$ via an ODMR measurement. The effective two-level system we use is formed by the sublevel $\ket{m_s=-1}\equiv \ket{\uparrow} $ and $\ket{m_s=0} \equiv \ket{\downarrow}$. Interaction with $^{13}$C nuclear spin bath is the dominant source of decoherence for ultrapure IIa type crystals \cite{Rein12} which were used in our experiments. The magnetic noise can be modelled by a fluctuating field and its spectrum is expected to be Lorentzian \cite{Hanson08,Rein12}. While significant progress was achieved in material engineering \cite{Bal09} it will not be possible to eliminate all sources of magnetic noise in this way. This is in particular so for implanted NV defects \cite{Neumann10}. Thus this relatively slow bath is both practically important and at the same time ideal for decoupling experiments \cite{Nay11} and thus for the demonstration of CCD.

%This relatively slow bath is ideal for decoupling experiments \cite{Nay11}. Significant progress was achieved in material engineering %\cite{Bal09} but not all sources of magnetic noise can be eliminated in this way, in particular for implanted NV defects %\cite{Neumann10}.

In our experiment, we first polarize the NV center electron spin into the sublevel $m_s=0$ with a green laser (532 nm). We note that this is equivalent to the preparation of an initial coherent superposition of the dressed states, namely $\ket{m_s=0} \equiv \ket{\downarrow} = \sqrt{\frac{1}{2}} ( \ket{\uparrow}_x+\ket{\downarrow}_x)$. For comparison, we start by applying a single driving field on resonance with the electronic transition $m_s=0 \leftrightarrow m_s=-1$. We measure the oscillation of the state $m_s=0$ population. It can been seen from Fig.\ref{Rabi}(a) that the decay of the Rabi oscillation is very fast. Theoretically, we model the microwave fluctuation by an Ornstein-Uhlenbeck process, and thus the amplitude of the driving field is time-dependent with random fluctuation as $\Omega_i(t) =\Omega_i [1+\delta_i(t)]$. The system dynamics is described by the following master equation
\begin{equation}
\frac{d}{dt}\rho =-i[H(t),\rho]+ \frac{\Gamma}{2}\sum_{a=\sigma_\pm}(2a^{\dagger} \rho a -\rho a a^{\dagger} -a a^{\dagger}\rho)
\end{equation}
The relaxation takes the high temperature limit, which is the case for our experiment at room temperature. We choose the relaxation parameter $\Gamma$ corresponding to $\mbox{T}_1=1.5 \mbox{ms}$ which is close to the value of the diamond used in our experiment. The numerical result agrees well with the experimental data and fits a Gaussian decay envelope $S_1(\tau)=\exp{\bla{-b_1^2 \tau^2/2}}$ for slow fluctuating fields \cite{KuboBook}, see Fig.\ref{Rabi}(a).  We thus estimate that the decay rate of Rabi oscillation is $b_1 \approx 98 \mbox{kHz}$, and the decay time scale is $\tau_1=2.3 \mu s$ defined by $S(\tau_1)=e^{-1}$. We stress that the fast decay is mainly due to intensity fluctuations of the microwave field. To support this observation, we have simulated the decay of coherence for two different Rabi frequencies $\Omega_1=20$ and $40 \mbox{MHz}$. For comparison, we also perform simulations assuming that there is no microwave fluctuation. Our results are shown in Fig.\ref{Rabi}(a2) which provides evidence that the fast decay mainly stems from the microwave fluctuation, while the residual effect of magnetic noise from the slow spin bath formed by $^{13}$C nuclei is comparatively smaller.

%as the Rabi frequency is $40 \mbox{MHz}$, which is strong enough to suppress the effect of magnetic noise almost completely (the %nuclear spin bath formed by $^{13}$C nuclei is slow compared to driving frequency). This statement will be supported and verified by %the experimental data that we present here.

To demonstrate the working principle of our concatenated continuous dynamical decoupling, we add a second-order driving field with a weaker amplitude than the first driving \cite{footnote}. We first show that CCD scheme can sustain Rabi oscillations by applying a second-order driving field, the intensity of which is ten times weaker than the first driving field. In Fig.\ref{Rabi}(b), we observe coherent Rabi oscillation after $300, 500, 800,1000 \mu s$, and find that it decays significantly slower than the one using only a single driving field, see Fig.\ref{Rabi}(a) for comparison. Our experimental data thus demonstrates that the effect of the fluctuation of the first driving field can be significantly suppressed by a second-order driving field, see Appendix for more discussions.

\begin{figure}[t]
\begin{center}
\begin{minipage}{12cm}
\hspace{-0.6cm}
\includegraphics[width=6cm]{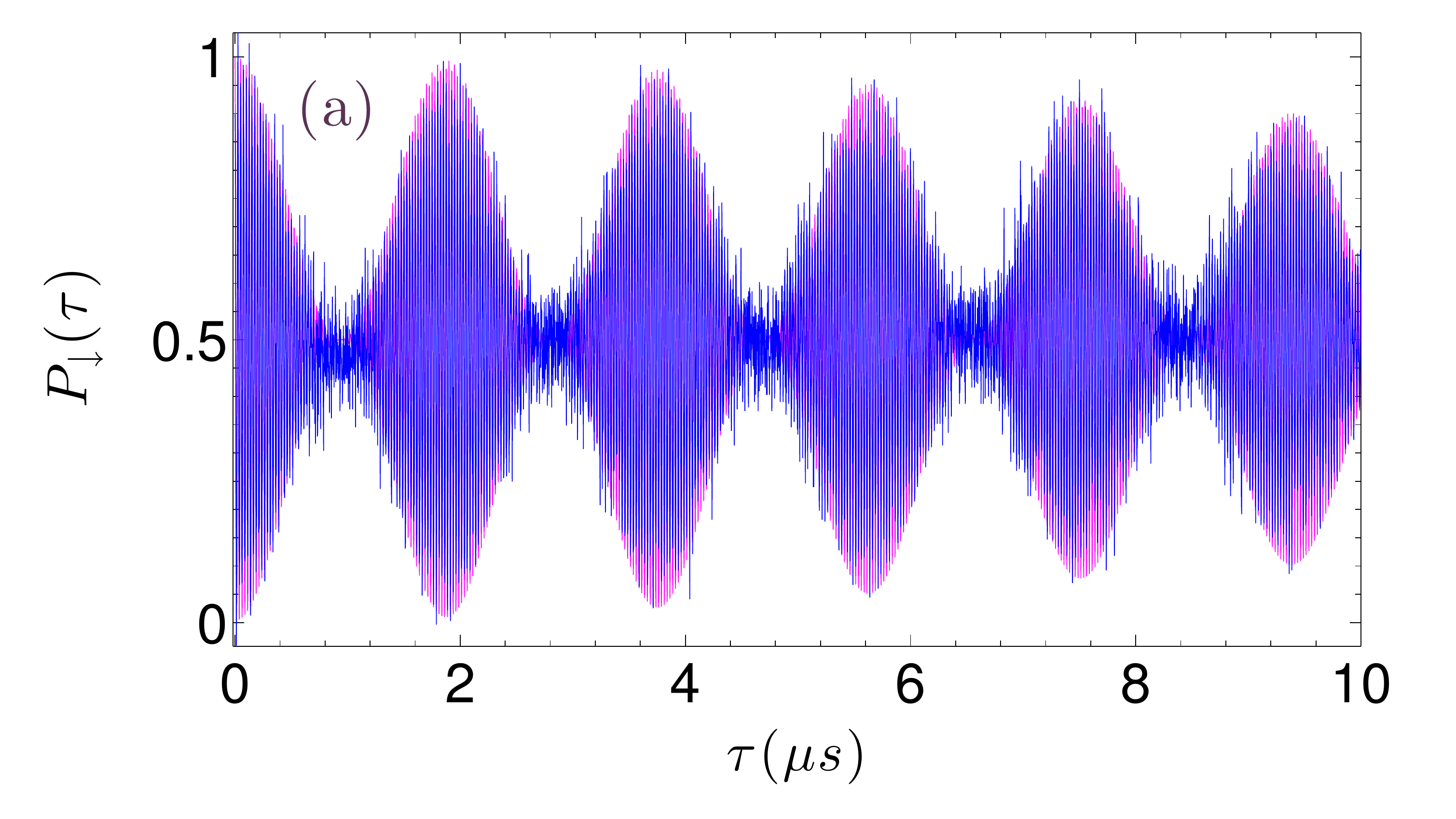}
\hspace{-0.4cm}
\includegraphics[width=6cm]{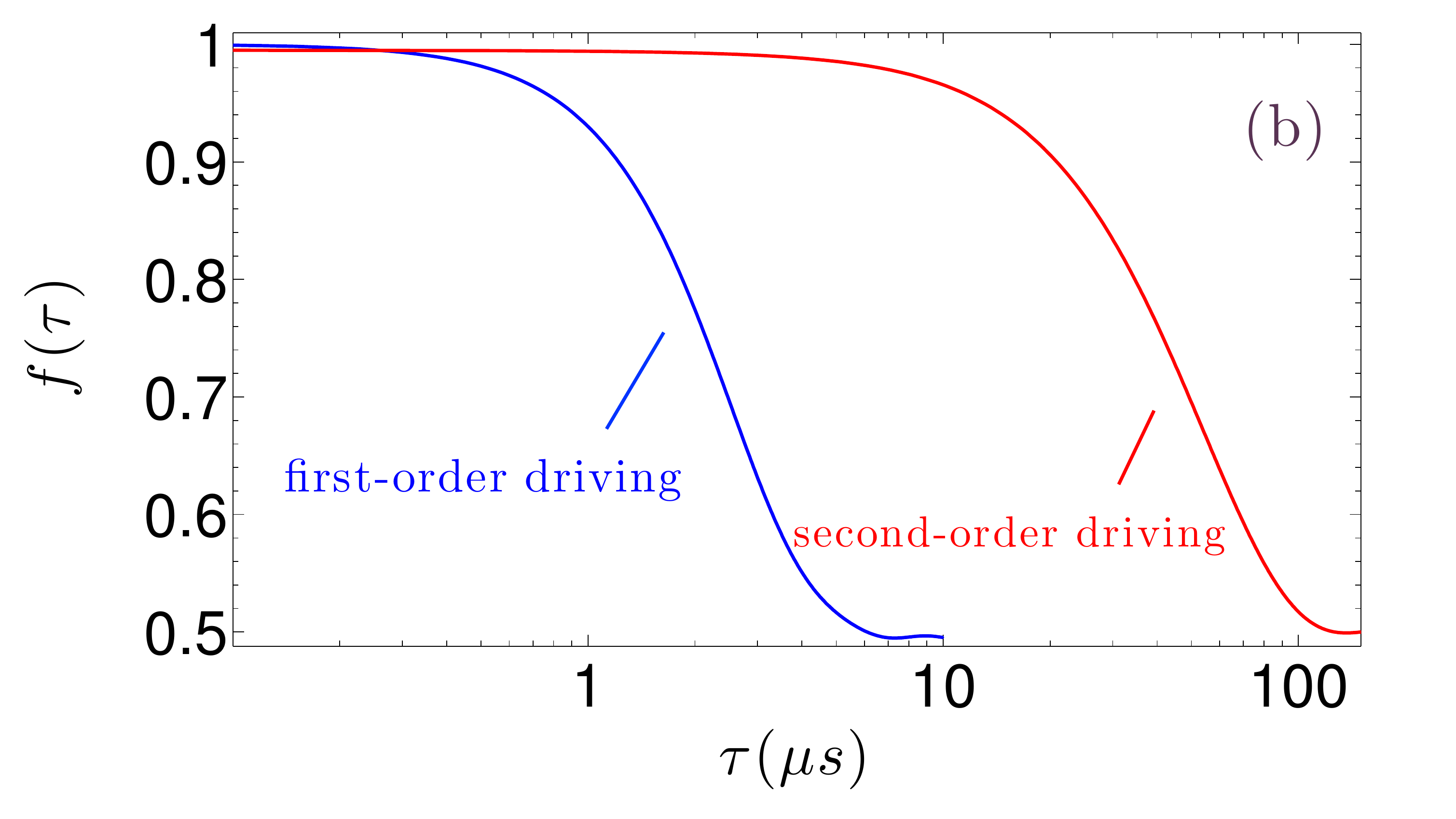}
\end{minipage}
\begin{minipage}{12cm}
\hspace{-0.6cm}
\includegraphics[width=6cm]{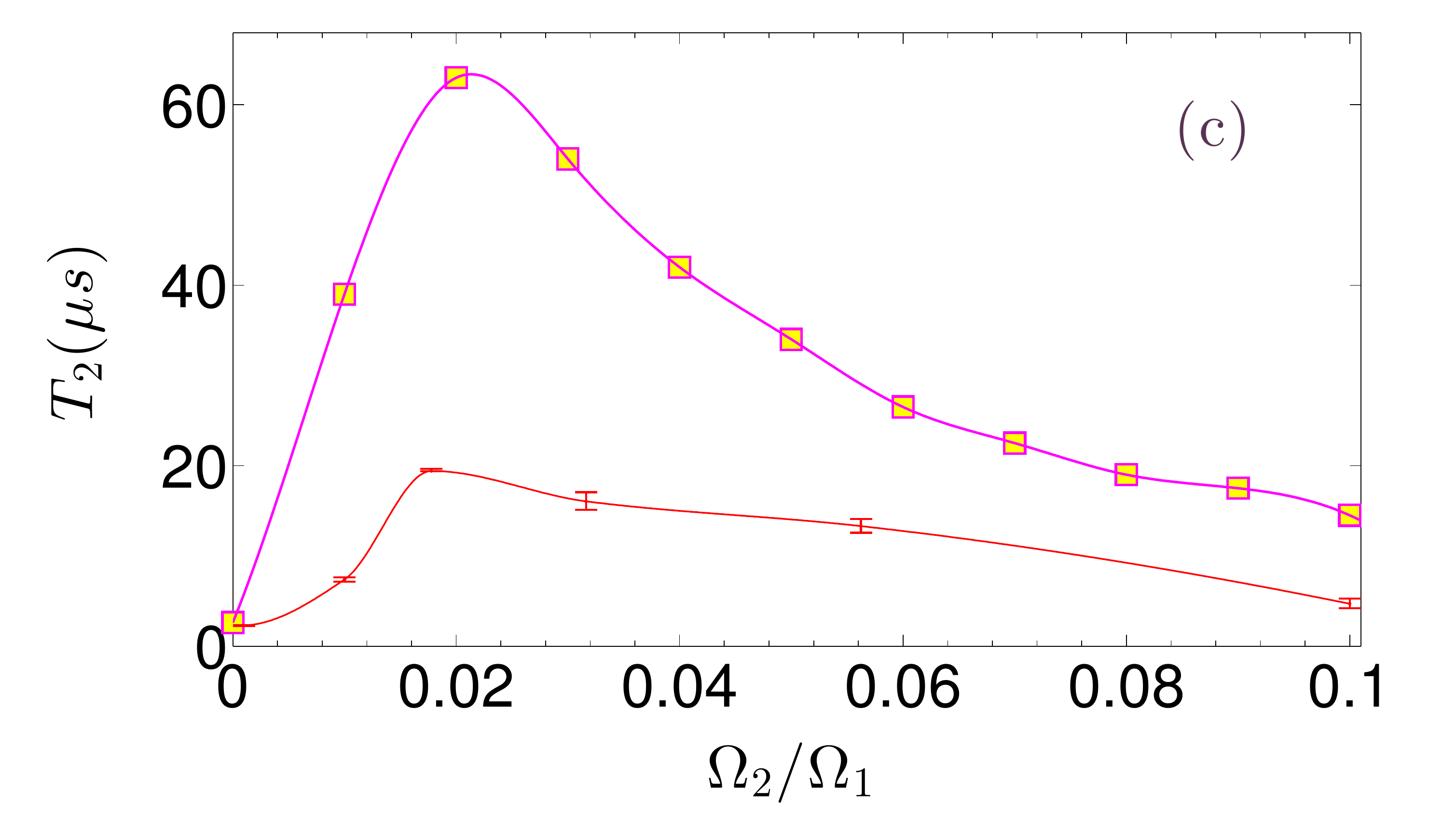}
\hspace{-0.4cm}
\includegraphics[width=6cm]{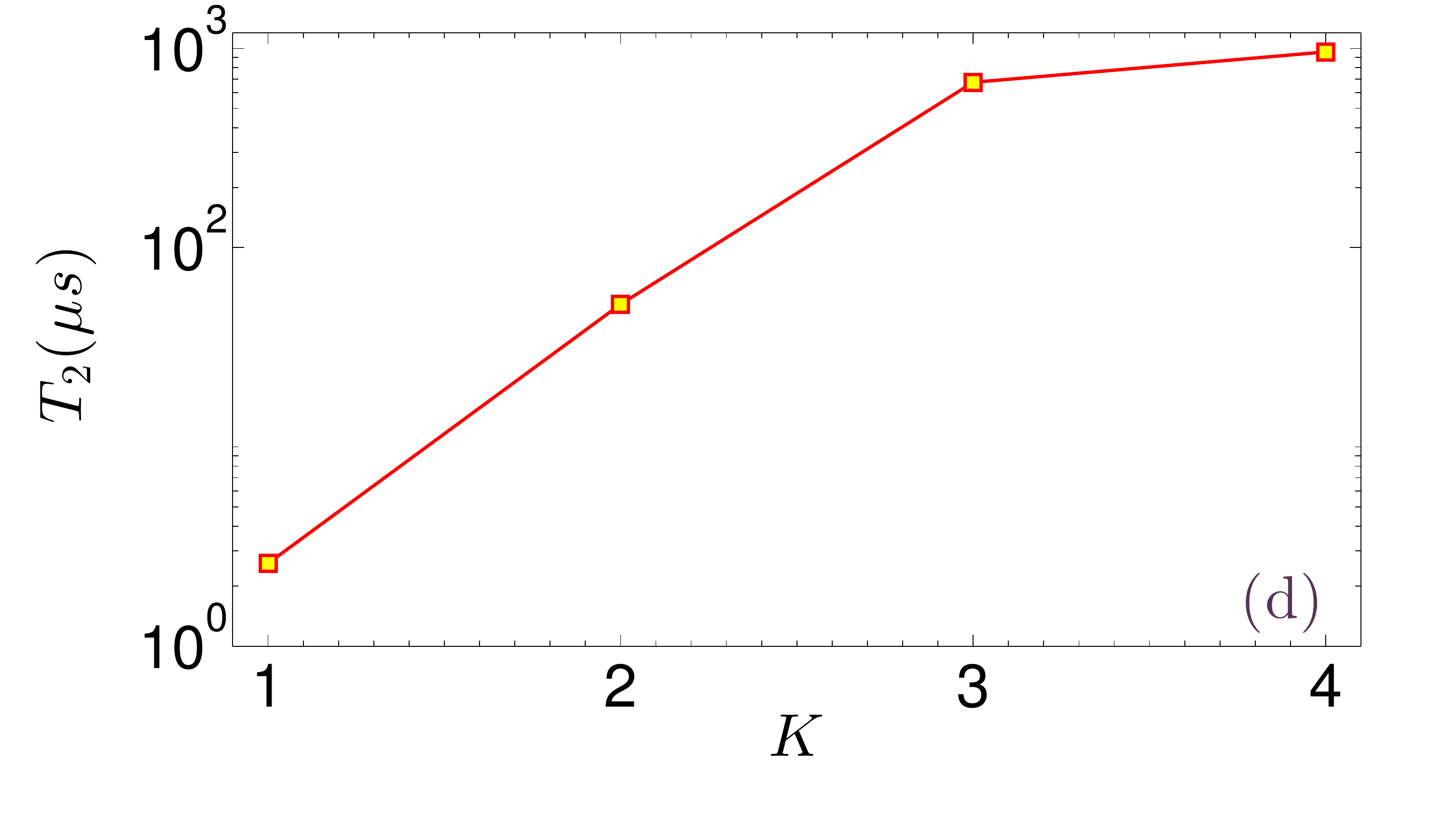}
\end{minipage}
\end{center}
\caption{(Color online) \textbf{a}: The measured state population of the sublevel $m_s=0$ as a function of the evolution time $\tau$. The blue curve is the experimental data, and the purple curve is numerical simulation using the parameters estimated from the Rabi experiment of single driving. The Rabi frequency of the first driving is $\Omega_1=40\mbox{MHz}$, and the intensity of the second driving is three orders of magnitude weaker. \textbf{b}: Numerical simulation for the decay of coherence using first- and second-order driving scheme using the field in Eq.(\ref{Hd2-m}). \textbf{c}: The coherence times extracted from our experimental data (blue, diamond) of the second-order dressed qubit as a function of the amplitude of the second-order driving, and the further improved coherence times  (purple, square) calculated from numerical simulation with a refined second-order field in Eq.(\ref{Hd2-m}) as a function of the amplitude of the second-order driving. The error bars are obtained with above $95\%$ confidence. \textbf{d}: The scaling of the coherence time with $K$th-order driving fields, see Eq.(\ref{Hd2k1}-\ref{Hd2k}). The relaxation time is $\mbox{T}_1=1.5\mbox{ms}$. In (b-d), the parameters are $\Omega_1=40\mbox{MHz}$ and $\Omega_{K}=\Omega_{K-1}/30$ for $K=2,3,4$.
% We fit the experimental data with a Gaussian decay envelop for the Rabi oscillation and the beatings from the second-order driving %respectively.
}\label{Ramsey}
\end{figure}

%\begin{figure}[b]
%\begin{center}
%\hspace{-0.5cm}
%\includegraphics[width=8.2cm]{FixPhase.pdf}
%\end{center}
%\caption{(Color online) The measured state population of the sublevel $m_s=0$ as a function of the evolution time $\tau$. The blue %curve is the experimental data, and the red curve is from numerical simulation with the estimated relative phase $\varphi=2\pi/5$. %The Rabi frequency of the first driving is $\Omega_1=30\mbox{MHz}$, and the second driving is $\Omega_2=\Omega_1/10$. %}\label{Fixedphase}
%\end{figure}

%We further estimate the dephasing time of the second-order dressed qubit by numerically fitting the decay of beatings (i.e. the third %term in Eq.(6)), which reflects the coherence of the second-order dressed qubit. The decay stems from the fluctuations of both the %first and second driving fields. Our experimental data, in combination with numerical fitting, as plotted in Fig.3 (b), shows that there is %an optimal amplitude for the second-order driving field in order to achieve the best coherence time.

We further demonstrate a significant improvement of the dephasing time. In our experiment, the concatenated driving fields are generated by using arbitrary waveform generator (AWG). With a second-order driving field about $10^3$ times less intense than the first-order driving, we observed a prolongation of the coherence time by an order of magnitude from $\mbox{T}_2=2.3\mu s$ to $\mbox{T}_2\cong 21\mu s$ \cite{footnote} as estimated by numerical fitting the decay of beat oscillation, see Fig.\ref{Ramsey}(a). We stress that such a weak additional driving significantly improves the coherence time mainly by decoupling the amplitude fluctuation of the first driving field (and incidentally further suppressing the residual effect of magnetic noise). By increasing the amplitude of the second-order driving one can suppress the effect of the fluctuation in the first driving field more effectively, however this will result in larger fluctuations of the second-order driving itself. The compromise between these two effects leads to the optimal choice of a second-order driving. This observation has also been confirmed by our experiment data, from which we estimate the coherence times for various strength of the second-order driving field, see Fig.\ref{Ramsey}(c) (blue, diamonds). In the experiment we have used the simplified second-order field $H_{d_2}=\hbar\Omega_2  \cos{\blb{\bla{\omega+\Omega_1}t+\varphi}}\sigma_x$ for which some counter-rotating terms persist that limit the achievable coherence time. This limitation can be overcome by using the refined second-order field given in Eq.(\ref{Hd2-m}) as confirmed by our numerical simulation shown in Fig.\ref{Ramsey}(b) and Fig.\ref{Ramsey}(c) (red squares). These simulations use the parameters estimated from the experiment and exhibit the same qualitative features as the measured data shown in Fig.\ref{Ramsey}(c) (blue, diamonds). To further extend $\mbox{T}_2$, one would need to apply higher-orders driving fields. Our calculations show that, see Fig.\ref{Ramsey}(d), the coherence time can approach to the $\mbox{T}_1 $ limit ($\mbox{T}_1=1.5 \mbox{ms}$ in the present example) with the fourth-order scheme (i.e. adding the driving fields as in Eq.(\ref{Hd2k1}-\ref{Hd2k})) for the current experiment parameters. The general ability of high-order schemes to suppress decoherence from the environment {\em and} the driving fields is valid on condition that the microwave fluctuation is slow (i.e. its bandwidth is smaller $1/T_1$) which is usually the case in experiments. This implies that the width of the microwave fluctuation spectrum (which is proportional to the inverse of the correlation time of fluctuation) is relatively small. Thus, high-order driving fields can still effectively suppress the effect of the proceeding field fluctuation as long as their amplitude is much larger than the width of the microwave noise spectrum.

%general ability When higher orders are added, are such
%remarkable increases in coherence time extended? What is the scaling
%to the T1 limit in terms of ‘n-th’ orders of CCD?

%In our experiment, with a second-order driving field about 1000 times less intense than the first-order driving, one can prolong the %coherence by an order of magnitude from $T_2=2.5\mu s$ to $T_2\cong 25\mu s$, see Fig.3 (a-b). Such a weak additional driving %would not have apparent effect in suppressing environmental magnetic noise, nevertheless it significantly improves the dynamical %decoupling efficiency. This fact supports our claim that the fast decay of Rabi oscillation in Fig.2(a) is mainly due to the fluctuation in %the driving field, which is shown to be a severe problem in conventional continuous dynamical decoupling with a single drive.

%We also demonstrate another important ingredient for concatenated continuous dynamical decoupling, namely fixing the relative %phase between consecutive driving fields, in order to generate a long living dressed qubit. We achieved this goal by using arbitrary %waveform generator (AWG) to generate the first- and second-order driving fields. In Fig.4, we show an example measurement with an %estimated fixed phase $\varphi=2\pi/5$.

\section{Coherent manipulation of a dressed qubit in CCD scheme}
\label{sec:man}

\begin{figure}[b]
\begin{center}
\hspace{0cm}
\includegraphics[width=7cm]{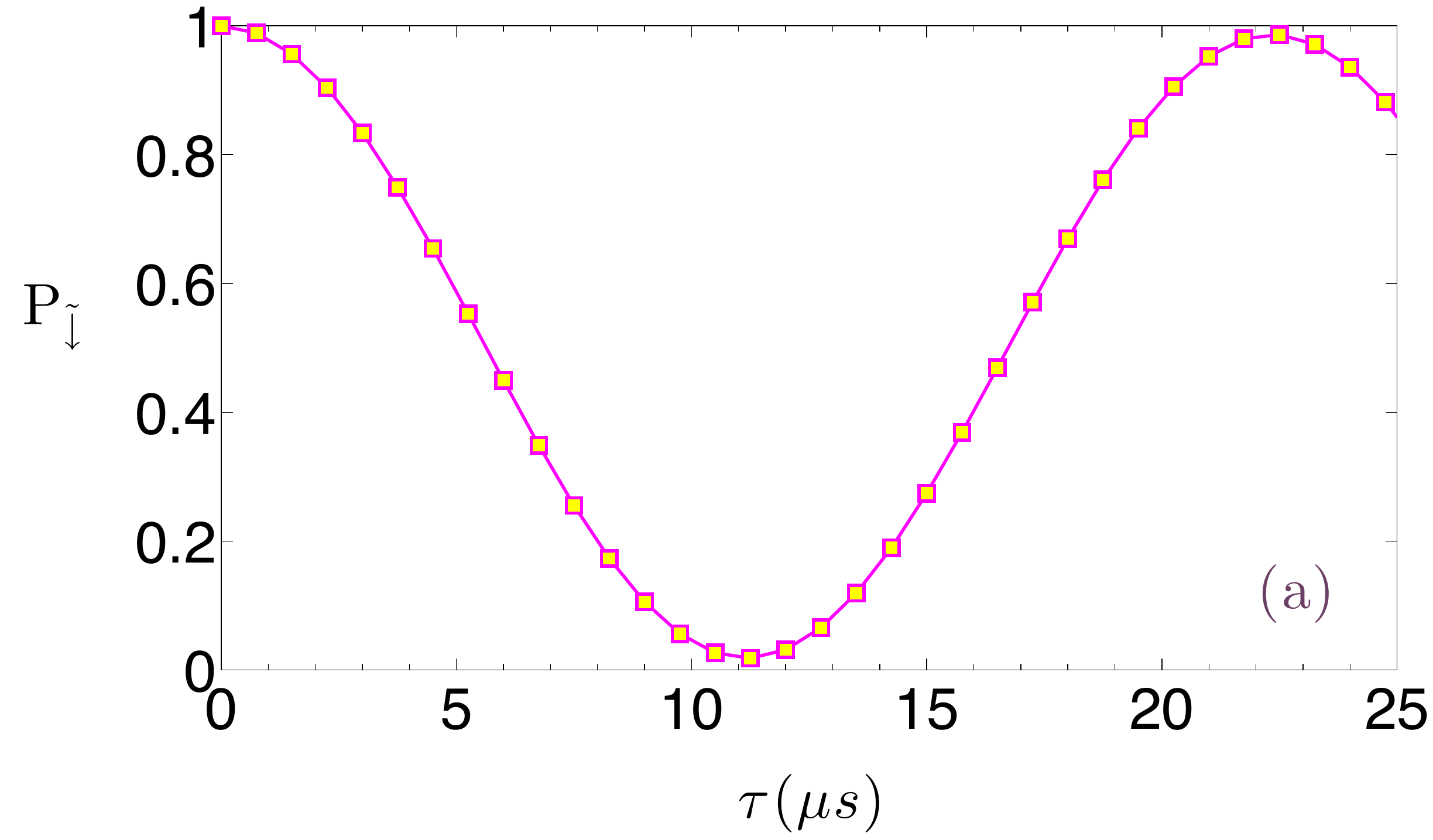}
\hspace{1cm}
\includegraphics[width=6.5cm]{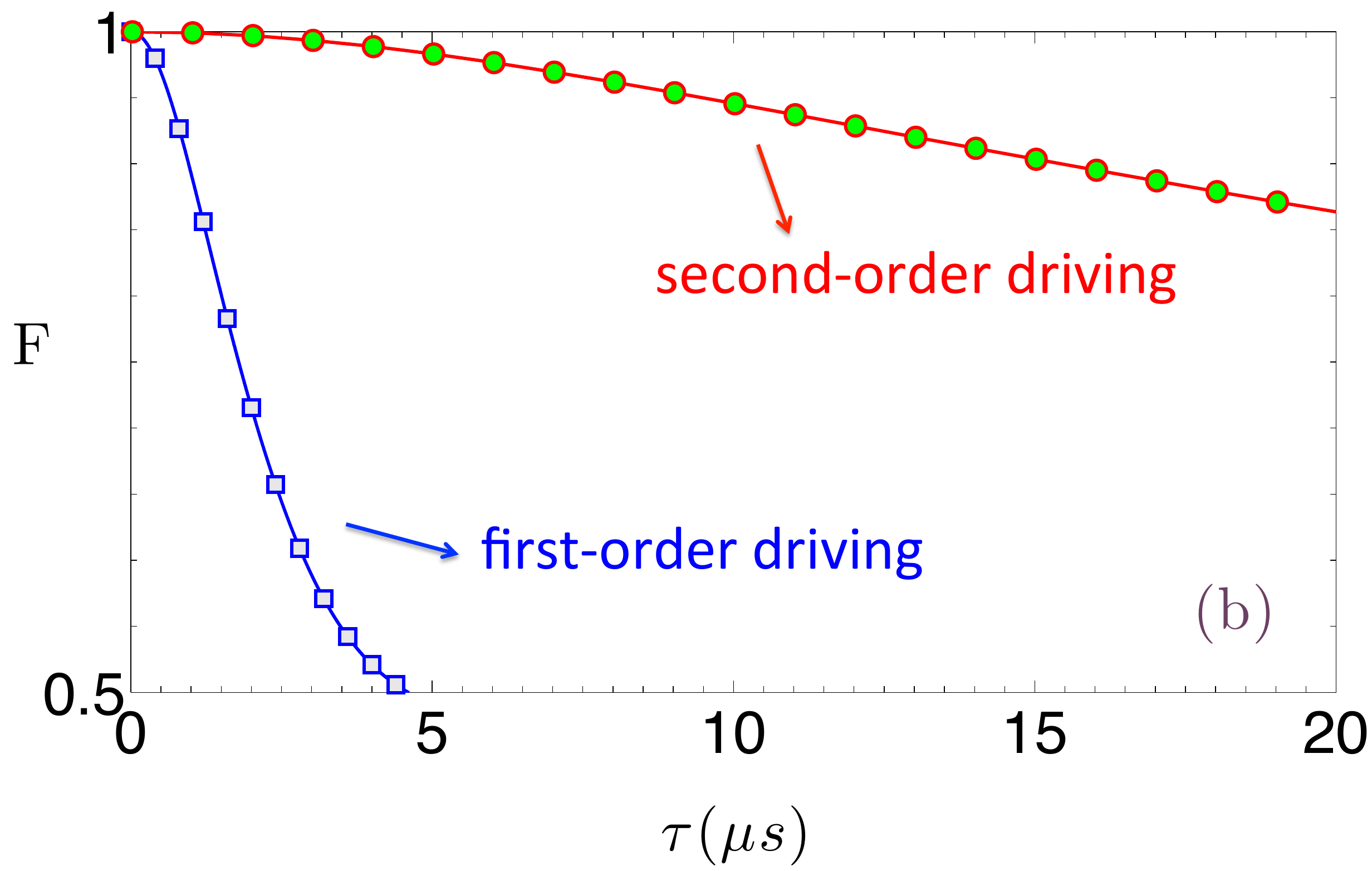}
\end{center}
\caption{(Color online) (a) Coherent manipulation of a second-order dressed qubit: the population of the dressed state $\ket{\tilde{\downarrow}}\equiv \ket{\downarrow}_y$ as a function of time $\tau$. The parameters are $\Omega_1=40\mbox{MHz}$, $\Omega_2=\Omega_1/30$ and $\Omega=\Omega_2/15$, $\phi_2=0$, see Eq.(\ref{Hsg-s}). (b) The fidelity of two-qubit coupling with respect to the corresponding ideal (noiseless) effective coupling Hamiltonian. The dipole-dipole interaction strength is $J=50 \mbox{kHz}$. The other parameters are $\Omega_1=40\mbox{MHz}$, $\Omega_2=\Omega_1/30$.}\label{Qgate:s1}
\end{figure}

Dressed qubits can be coherently manipulated by external driving fields and coupled with each other via electron spin dipole interaction. For single qubit rotation, we can use the following radio-frequency field
\begin{equation}
H_{sg}=\hbar\Omega \left[\prod_{k=1,\cdots, n} \cos(\Omega_k t+\varphi_k) \right] \sigma_z.
\label{Hsg-s}
\end{equation}
To demonstrate explicitly how this works, we consider the second-order dressed qubit as an example and apply the field $H_{sg}=\hbar\Omega  \cos(\Omega_1 t)\cos(\Omega_2 t+ \varphi_2)  \sigma_z
$. In the first-order interaction picture with respect to the original Hamiltonian $H_0^{(1)}=\frac{\hbar\omega}{2} \sigma_z$, the coupling can be written as
\begin{equation}
H_c^{(1)}= \frac{\hbar\Omega_1 }{2} \sigma_x + \hbar\Omega_2   \cos(\Omega_1 t) \sigma_y+\hbar\Omega  \cos(\Omega_1 t)\cos(\Omega_2 t+ \varphi_2)\sigma_z.
\end{equation}
In the second-order interaction picture with respect to $H_0^{(2)}=\frac{\hbar\Omega_1}{2} \sigma_x$, and using the rotating wave approximation, it becomes
\begin{equation}
H_c^{(2)}=\frac{\hbar\Omega_2 }{2} \sigma_y + \frac{\hbar\Omega}{2} \cos{\bla{ \Omega_2 t + \varphi_2}} \sigma_z.
\label{Hc2-s}
\end{equation}
By choosing appropriate phase $\varphi_2$, one can implement general rotations of the second-order dressed qubit (encoded in the eigenstates of $\sigma_y$: $\ket{\uparrow}_y$ and $\ket{\downarrow}_y$), e.g. see Fig.\ref{Qgate:s1}(a). Similar results can be obtained for general higher-order dressed qubits. In fact, this also implies the possibility to measure the amplitude of the RF-filed $\Omega$  by Rabi-type spectroscopy with the robust dressed qubit, e.g. see Eq(\ref{Hc2-s}). The sensity is determined by the coherence time of the dressed qubit \cite{Taylor08,Fedder10}. Using concatedated continuous dynamical decoupling, the coherence time of the dressed qubit is possible to be prolonged to the relaxation time, and one could thus in principle construct a magnetometer with the sensitivity scaling with  $1/\sqrt{\mbox{T}_1}$. We would also like to point out that CCD can also be combined with the sensing protocols in biological environments proposed in \cite{DS11} where the noisy biological environment requires strong driving. With the robust energy gap provided by CCD scheme, the Hartmann-Hahn condition \cite{Hahn62} will be more stably matched, and thus the measurement of the coupling between NV center and the target nuclear spin will be more efficient and precise.  \\

\section{Coherent coupling between CCD dressed qubits}
\label{sec:coupling}

Regarding the coupling between dressed qubits, we again take the second-order decoupling scheme as an example. The dipole-dipole interaction, e.g. between NV electron spins, is described by
\begin{equation}
H_{d-d}=J(3S_z^a S_z^b-\vec{S}_a\cdot \vec{S}_b).
\end{equation}
where $S$ is the spin-1 operator of NV centers. In the first-order interaction picture, the electron-electron interaction with the driving fields becomes \cite{Ber11}
\begin{equation}
H_{d-d}^{(1)}=2J S_z^a S_z^a + \hbar \Omega_1^a \sigma_x^a+ \hbar \Omega_1^b \sigma_x^b+\hbar\Omega_2^a \cos(\Omega_1^a t) \sigma_y^a+\hbar\Omega_2^b \cos(\Omega_1^b t) \sigma_y^b.
\end{equation}
We use the two sublevels $\ket{m_s=-1}\equiv \ket{\uparrow} $ and $\ket{m_s=0} \equiv \ket{\downarrow}$ of NV electron spin as a qubit. The above interaction in the qubit Hilbert subspace can be written as
\begin{equation}
H_{d-d}^{(1)}=\frac{J}{2} \bla{\sigma_z^a \sigma_z^b+\sigma_z^a +\sigma_z^b }+ \frac{\hbar \Omega_1^a}{2} \sigma_x^a+ \frac{\hbar \Omega_1^b}{2}  \sigma_x^b+\hbar\Omega_2^a \cos(\Omega_1^a t) \sigma_y^a+\hbar\Omega_2^b \cos(\Omega_1^b t) \sigma_y^b,
\end{equation}
where $\sigma_x$, $ \sigma_y$ and $\sigma_z$ are Pauli operators in the spin up and down basis.  The effective Hamiltonian in the second-order interaction picture with respect to $H_0^{(d)}=\hbar \Omega_1^a \sigma_x^a+ \hbar \Omega_1^b \sigma_x^b$ is
\begin{equation}
H_{d-d}^{(2)}=\frac{J}{4}\bla{ \sigma^a_y  \sigma^b_y + \sigma^a_z \sigma^b_z} +\frac{\hbar\Omega_2 ^a}{2}   \sigma_y^a+\frac{\hbar\Omega_2 ^b}{2}   \sigma_y^b.
\label{qcoupling-s}
\end{equation}
The effective entangling coupling between dressed qubits is thus feasible (as in Eq.(\ref{qcoupling-s})), and can be exploited to construct two-qubit gates \cite{DoddSI}. We use the following quantity to characterize the fidelity \cite{Cabrera07} between the coherent coupling of two dressed qubits and the evolution ($U_{\mbox{ideal}}$) from the ideal (noiseless) effective coupling Hamiltonian
\be
\mbox{F}\bla{U_{\mbox{ideal}},\mathcal{M}}=\frac{1}{16}\blb{4+\frac{1}{5}\sum_{\mu,\nu}\mbox{tr}\blb{\bla{U_{\mbox{ideal}}\bla{\sigma_{\mu}\otimes \sigma_{\nu}} U^{\dagger}_{\mbox{ideal}} }\cdot \mathcal{M}\bla{\sigma_{\mu}\otimes \sigma_{\nu}} }}
\ee
where $\sigma_{\mu}, \sigma_{\nu} =\mathbf{I},\sigma_x,\sigma_y,\sigma_z$ and $\bla{\sigma_{\mu}\otimes \sigma_{\nu}}\neq \mathbf{I}\otimes \mathbf{I}$. We numerically simulate the dynamics $\mathcal{M}$ by solving the mater equation in Eq.(\ref{ME-s}) including both the external magnetic noise and microwave field fluctuations. The result is shown in Fig.\ref{Qgate:s1}(b), which demonstrate that CCD scheme can also be helpful in the protection of two-qubit coupling.

\section{Potential applications of CCD schemes}

The CCD scheme can be combined with various quantum information processing tasks and make them robust against not only noise but also the fluctuations of decoupling fields. The coherent manipulation of the n-th order dressed qubit can be implemented with a radio frequency field, see section \ref{sec:man}. Such a qubit encoded with the sequential dressed states has an ultralong coherence time limited only by its $T_1$ time, and thereby can be exploited to construct a single-spin magnetometer \cite{Taylor08} to probe a weak oscillating magnetic field with an improved sensitivity.
%$b(t)=b_{AC}\cos[(\Omega_1+\Omega/2)t]$. If $b_{AC}$ is much larger than the fluctuation of the second driving field, the sensitivity %will be $\sim 1/\sqrt{T_1}$ with $T_1$ the coherence time demonstrated in Fig.2(a); otherwise it scales as $1/\sqrt{T_2}$ with $T_2$ %the prolonged dephasing time.
The scheme can also be beneficial for the construction of a precise noise spectrometer \cite{Yuge11,Byl11,Suter11b}. We note that the inhomogeneity over an ensemble of quantum systems (namely spatial fluctuation) leads to dephasing of the ensemble collective state.
However, if the amplitudes of the first order driving field exceeds the disorder, then the the effect of spatial fluctuations is suppressed (see e.g. \cite{QCED} for a detailed discussion). This again points to the usefulness of CCD as it allows us to increase the driving field amplitudes without suffering from the concomitant noise due to amplitude fluctuations.

%Our schemes can suppress this kind of spatial fluctuation and in the mean time is robust against the inhomogeneity of the driving field %itself. We expect that this may be useful in the ensemble-based magnetometer and quantum memory \cite{QCED}.

\section{Summary}

We have introduced the concept of CCD and implemented it experimentally. Using an NV center in diamond we demonstrated the superior performance of concatenated continuous dynamical decoupling compared to single driving fields in extending coherence times of a dressed qubit. A qubit encoded in the concatenated dressed states is robust against both environmental dephasing noise and intensity fluctuation of driving fields. Our schemes can be applied to a wide variety of quantum systems where they may find applications in the construction of nano-scale magnetometry and imaging e.g. with the NV center in diamond, and in the construction of fault-tolerant quantum gates that are protected against noise and control errors.

\section{Acknowledgements}

The work was supported by the Alexander von Humboldt Foundation, the EU Integrating Project Q-ESSENCE, the EU STREP PICC and DIAMANT, the BMBF Verbundprojekt QuOReP, DFG (FOR 1482, FOR 1493, SFB/TR 21) and DARPA. J.-M.C was supported also by a Marie-Curie Intra-European Fellowship within the 7th European Community Framework Program.

\section{Appendix}

\subsection{Driving fields for general concatenated continuous dynamical decoupling}

In our concatenated scheme, in general, we can apply n-th order continuous driving fields on the condition that two subsequent drivings describe rotations about orthogonal axes in the interaction picture (see the example for 2nd order driving fields for illustration). The original system Hamiltonian is
\begin{equation}
H_0=\frac{\hbar\omega}{2} \sigma_z + \frac{\hbar\delta_b(t)}{2} \sigma_z
\end{equation}
where $\sigma_z=\bla{\ketbra{\uparrow}{\uparrow}-\ketbra{\downarrow}{\downarrow}}$ and we have explicitly written the magnetic noise term $\delta_b(t)$. The first-order driving field (in the lab frame) is
\begin{equation}
H_{d_1}= \hbar\Omega_1 [1+\delta_1(t)]\cos{(\omega t)} \sigma_x
\label{Hd1-s}
\end{equation}
where $\delta_1(t)$ represents the amplitude fluctuation of the first order driving field. In the interaction picture with respect to $H_0^{(1)}=\frac{\hbar\omega}{2} \sigma_z$, we have the following effective Hamiltonian as
\begin{equation}
H_I^{(1)}=\frac{\hbar\Omega_1 }{2} [1+\delta_1(t)]\sigma_x + \frac{\hbar\delta_b(t)}{2} \sigma_z
\end{equation}
The first-order dressed states are the eigenstates of $\sigma_x$: $\ket{\uparrow}_x=\frac{1}{\sqrt{2}}\bla{\ket{\uparrow} +\ket{\downarrow}}$ and $\ket{\downarrow}_x=\frac{1}{\sqrt{2}}\bla{\ket{\uparrow} -\ket{\downarrow}}$. The effect of the magnetic noise term $\frac{\hbar\delta_b(t)}{2} \sigma_z$ will induce transitions between the dressed states $\ket{\uparrow}_x$ and $\ket{\downarrow}_x$, whose rate can be estimated by the noise spectrum of $\delta_b(t)$ at the transition frequency $\hbar\Omega_1$. The microwave fluctuation $\delta_1(t)$ causes the dephasing of the dressed states. To suppress the effect of fluctuations in the amplitude of the first-order driving, we apply a second-order driving field (in the labe frame) as
\begin{equation}
H_{d_2}=2\hbar \Omega_2[1+\delta_2(t)] \cos(\omega t+\frac{\pi}{2}) \cos(\Omega_1 t) \sigma_x.
\end{equation}
With such an additional driving field, the effective Hamiltonian with respect to $H_0^{(1)}=\frac{\hbar\omega}{2} \sigma_z$ (as in Eq.(\ref{Hd1-s})) becomes
\begin{equation}
H_I^{(1)}=\frac{\hbar\Omega_1 }{2}[1+\delta_1(t)] \sigma_x + \hbar\Omega_2 [1+\delta_2(t)] \cos(\Omega_1 t) \sigma_y+\frac{\hbar\delta_b(t)}{2} \sigma_z.
\end{equation}
In the second-order interaction picture with respect to $H_0^{(2)}=\frac{\hbar\Omega_1 }{2} \sigma_x$ we have
\begin{equation}
H_I^{(2)}=\frac{\hbar\Omega_2}{2}  [1+\delta_2(t)] \sigma_y + \frac{\hbar\Omega_1 }{2} \delta_1(t)\sigma_x
\label{HI2-s}
\end{equation}
where $\sigma_y=-i \ketbra{\uparrow}{\downarrow}+i \ketbra{\downarrow}{\uparrow}$ and the second-order dressed states are
$\ket{\uparrow}_y=\sqrt{\frac{1}{2}} \bla{\ket{\uparrow}+i\ket{\downarrow}}$ and  $\ket{\downarrow}_y=\sqrt{\frac{1}{2}} \bla{\ket{\uparrow}-i\ket{\downarrow}}$. For simplicity, here we only explicitly write the noise terms of $\delta_1(t)$ and $\delta_2(t)$ in Eq.(\ref{HI2-s}), nevertheless we stress that we do take it into account in our numerical simulations. Similarly, the fluctuation of the first driving field $\frac{\hbar\Omega_1 }{2} \delta_1(t)\sigma_x$ induces the transitions between the second-order dressed states $\ket{\uparrow}_y$ and $\ket{\downarrow}_y$. Its effect can be characterized by the power spectrum of $\delta_1(t)$ at the transition frequency $\hbar\Omega_2$. In the above derivations, we have adopted the rotating wave approximations (RWA) which hold when $\Omega_2 \ll \Omega_1 \ll \omega$, which are fulfilled in the present context. In a similar way, one can find the required driving fields for higher-order decoupling. For example, the third- and fourth-order driving fields can be provided by
\begin{equation}
H_{d_3} = 2\hbar\Omega_3 \cos(\omega t) \cos(\Omega_2 t) \sigma_x,\quad \mbox{and} \quad H_{d_4} = 4\hbar\Omega_4 \cos(\Omega_1 t )\cos(\Omega_3 t) \cos(\omega t+\frac{\pi}{2})\sigma_x.
\end{equation}
The general higher-order driving fields can be written in a similar way as in Eq.(4-5) of the main text.

\subsection{Numerical simulations of persistent Rabi oscillation}

\begin{figure}[t]
\begin{center}
\hspace{-0.2cm}
\includegraphics[width=4.5cm]{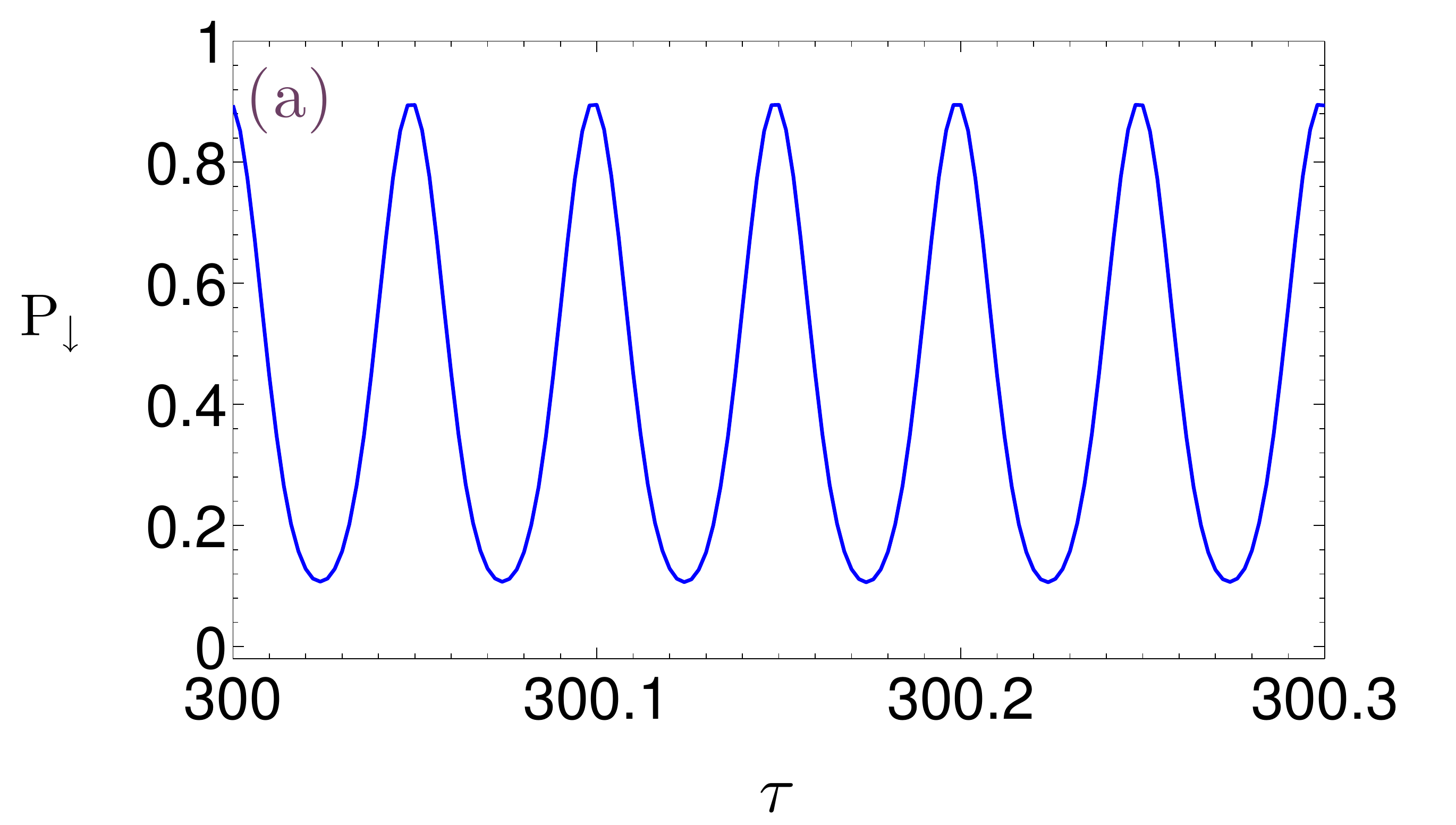}
\hspace{-0.3cm}
\includegraphics[width=4.5cm]{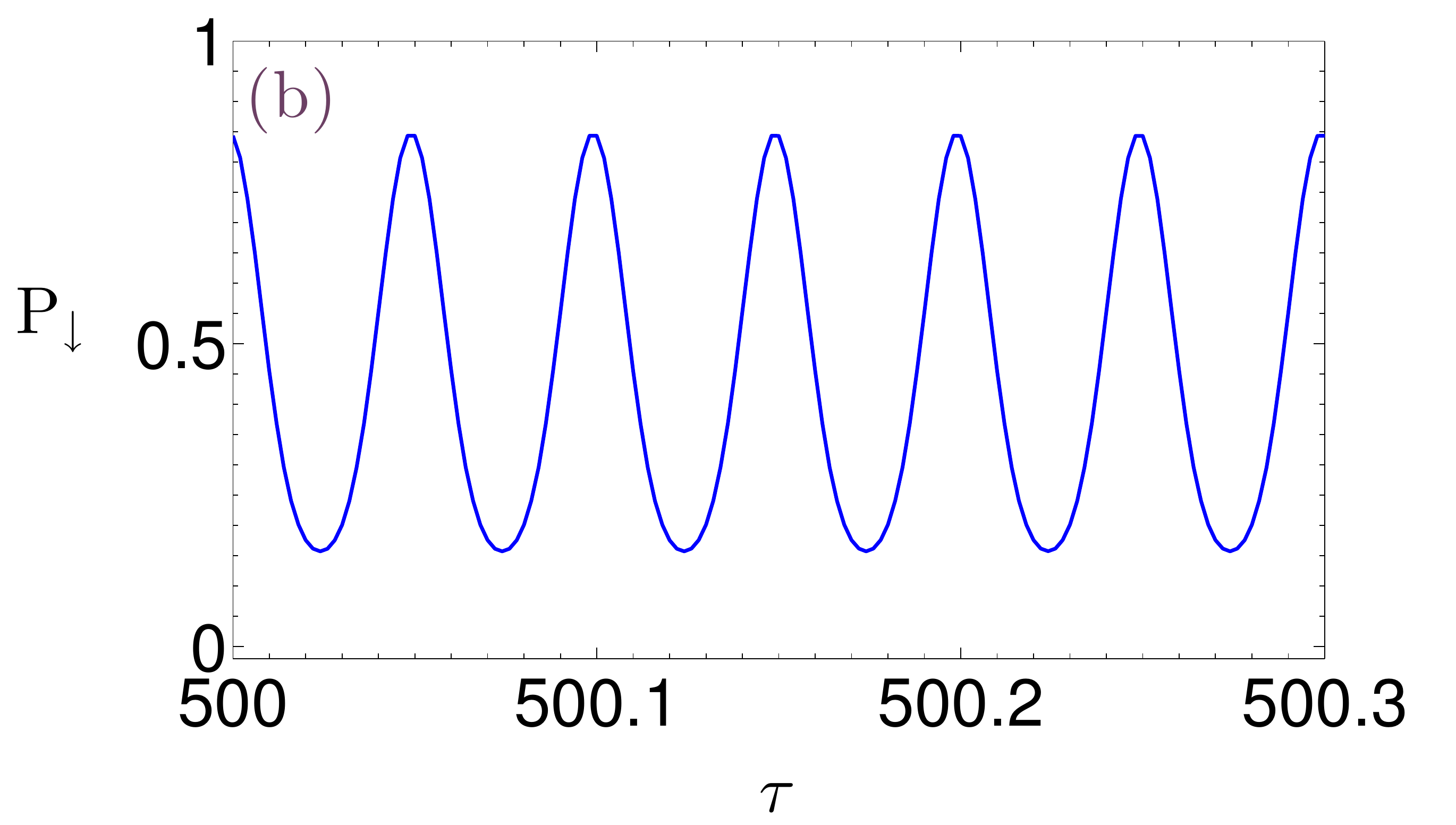}
\hspace{-0.3cm}
\includegraphics[width=4.5cm]{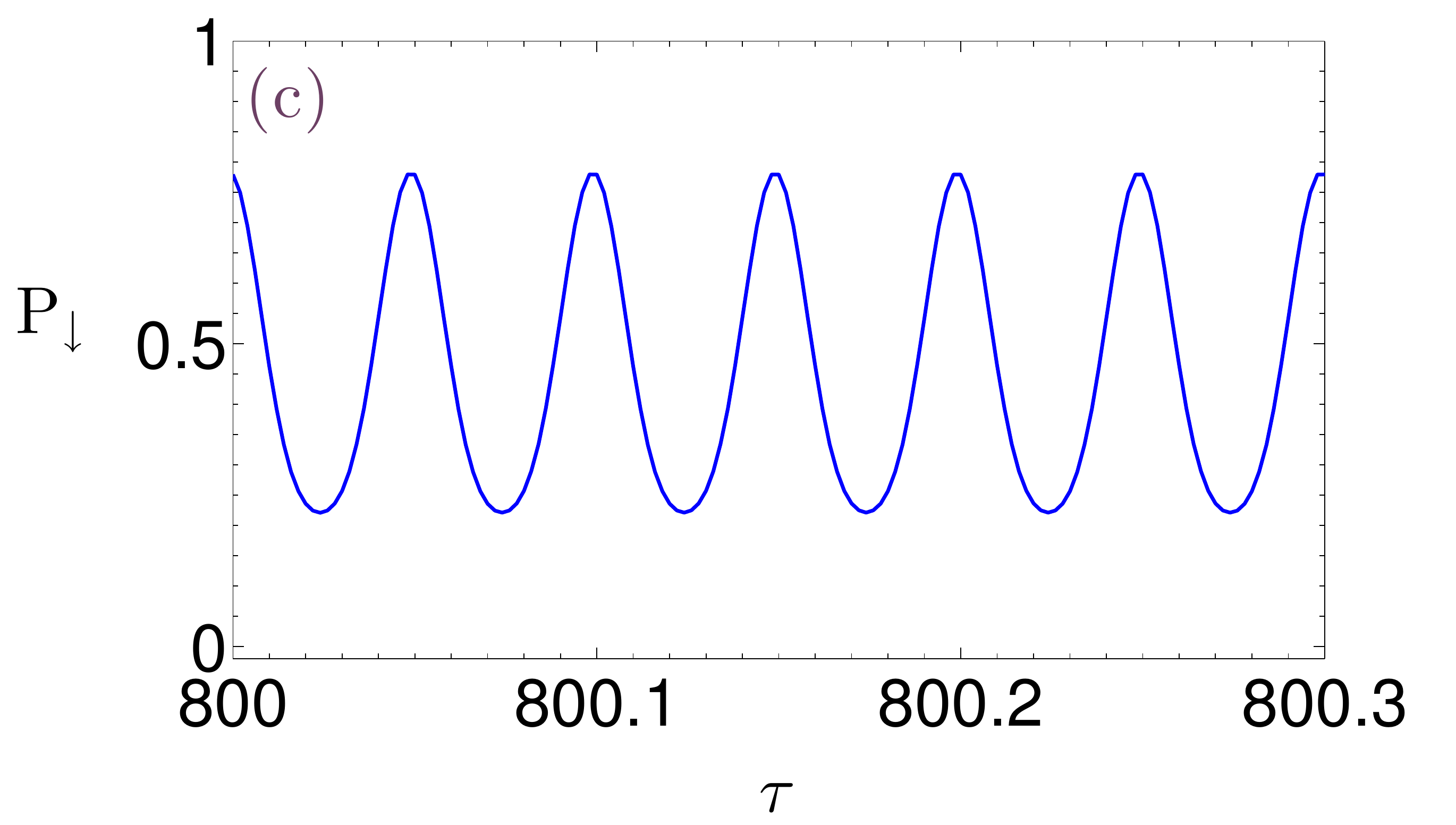}
\hspace{-0.3cm}
\includegraphics[width=4.5cm]{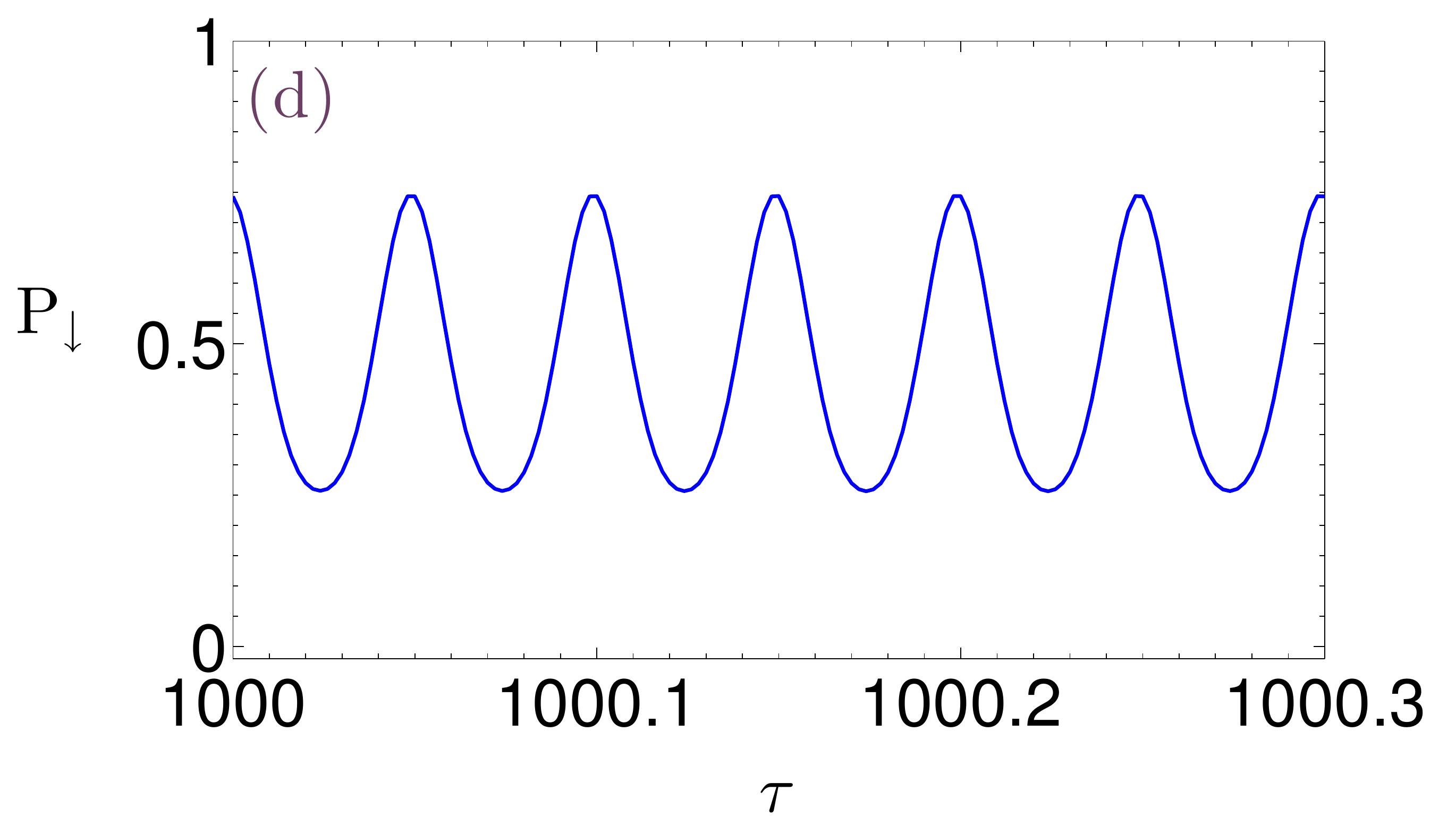}
\end{center}
\caption{(Color online) Numerical simulation of persistent Rabi oscillation by adding a second-order driving field, the intensity of which is 10 times weaker than the first driving ($\Omega_1=20 \mbox{MHz}$). The plot shows that Rabi oscillation can sustain for much longer time than one single driving field. The four panels show the very slow decay of the Rabi oscillation beyond $300 \mu s$. The decay is mainly due to the relaxation of NV center itself, we use the value of the relaxation time $T_1=1.5\mbox{ms}$ in the numerical simulations. The relative phase is fixed as $\varphi=0$.}\label{NSRabi}
\end{figure}

In our experiment, we apply a simplified second-order field as $H_{d_2}=\hbar\Omega_2  \cos{\blb{\bla{\omega+\Omega_1}t+\varphi}}\sigma_x$, which leads to the second-order effective Hamiltonian as
\begin{equation}
H_I^{(2)}=\frac{\hbar\Omega_2}{4} \bla{ \sin{\varphi}\sigma_y-\cos{\varphi}\sigma_z}
\end{equation}
We initially prepare NV spin in the state $\ket{\downarrow}\equiv \ket{m_s=0}$, which is one of the second-order dressed states if the relative phase $\varphi=0$. As we point out in the main text, the fluctuation of the first driving field would induce transitions of second-order dressed states, which can be suppressed due to the energy penalty induced by the second-order driving field. This is the working principle of CCD scheme. In Fig.2(a) of the main text, our experimental data demonstrates this principle by the observation of the persistent oscillation of the state $\ket{\downarrow}\equiv \ket{m_s=0}$ population. It represents the long lifetime of the second-order dressed states, which is dependent on the relaxation time $T_1$ of the bare spin states and the residual effect of the first driving field fluctuation. Our experimental data thus demonstrates that the effect of the first driving field fluctuation can be significantly suppressed by a second-order driving field. We have performed numerical simulation, as shown in Fig.\ref{NSRabi}, with a second-order driving field using the estimated parameters. We note that in the numerical simulation we fixed the relative phase, thus the contrast of Rabi oscillation is twice of Fig.2(b) in the main text. Nevertheless, the essential feature of the results from our numerical simulation, namely the slow decay of Rabi oscillation, agrees well with the experimental data.

\subsection{Numerical simulation of system dynamics with concatenated driving}

\begin{figure}[h]
\begin{center}
\begin{minipage}{20cm}
\hspace{-2.5cm}
\includegraphics[width=4.5cm]{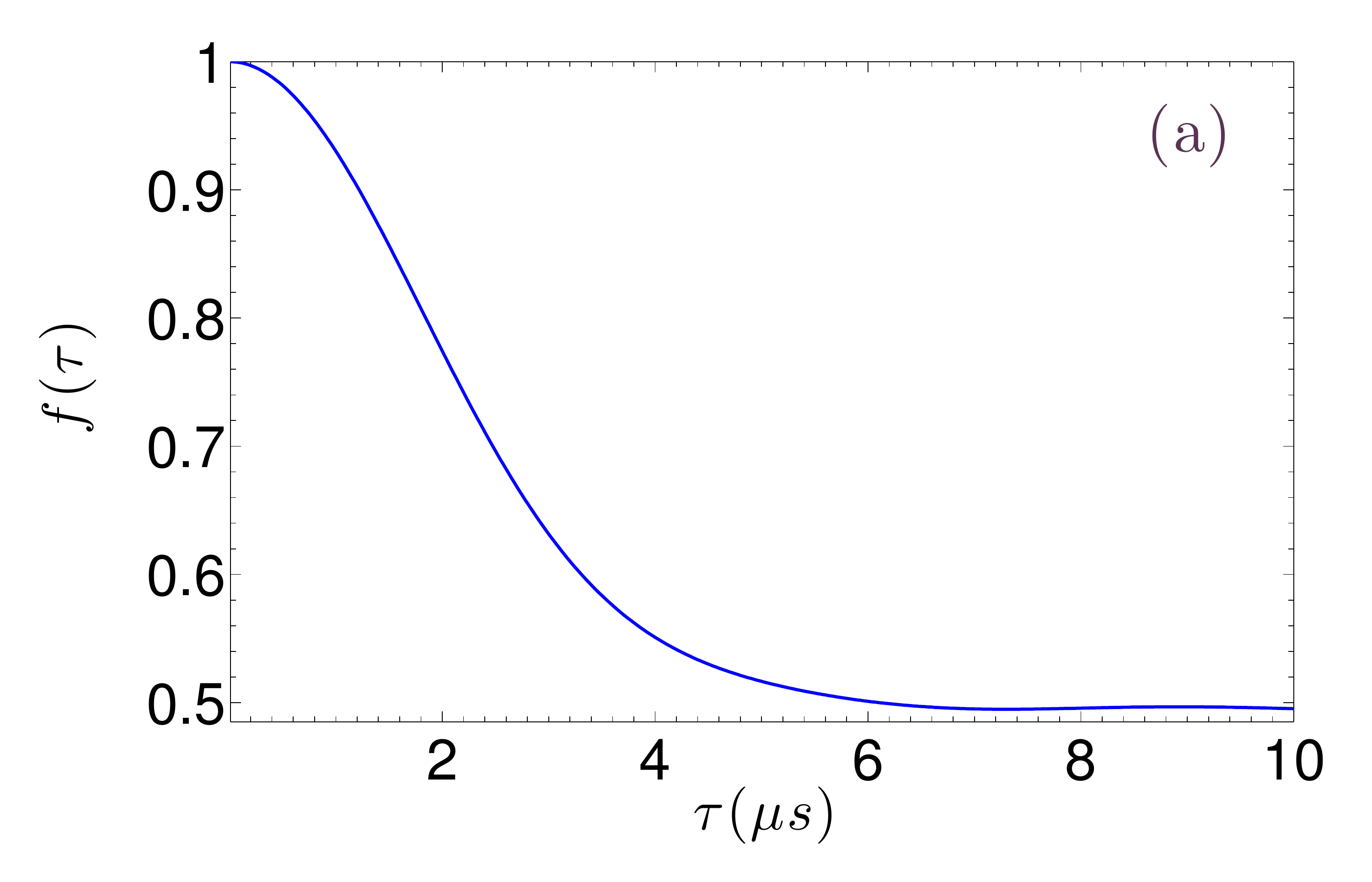}
\hspace{-0.2cm}
\includegraphics[width=4.5cm]{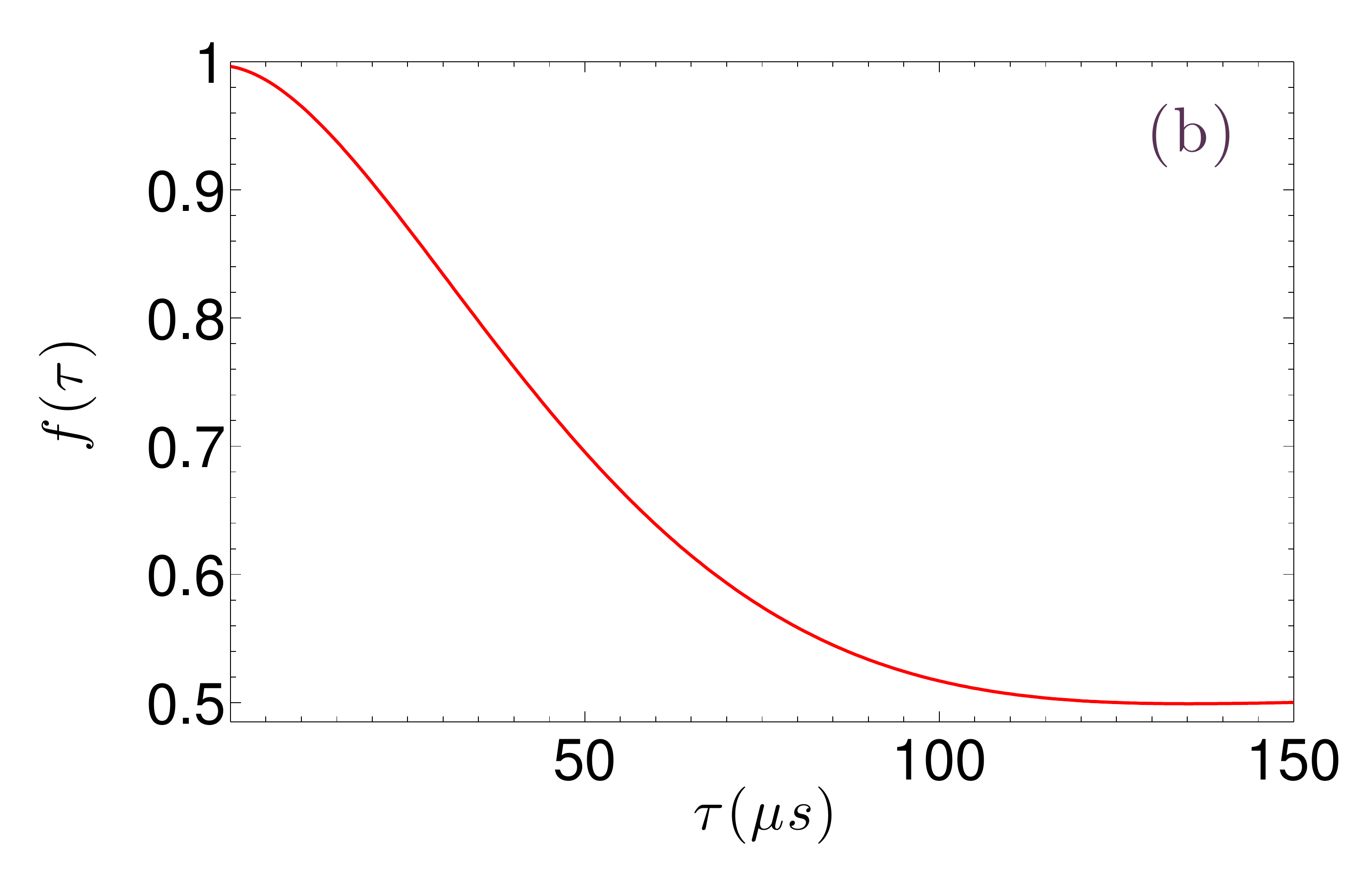}
%\end{minipage}
%\begin{minipage}{16cm}
\hspace{-0.2cm}
\includegraphics[width=4.5cm]{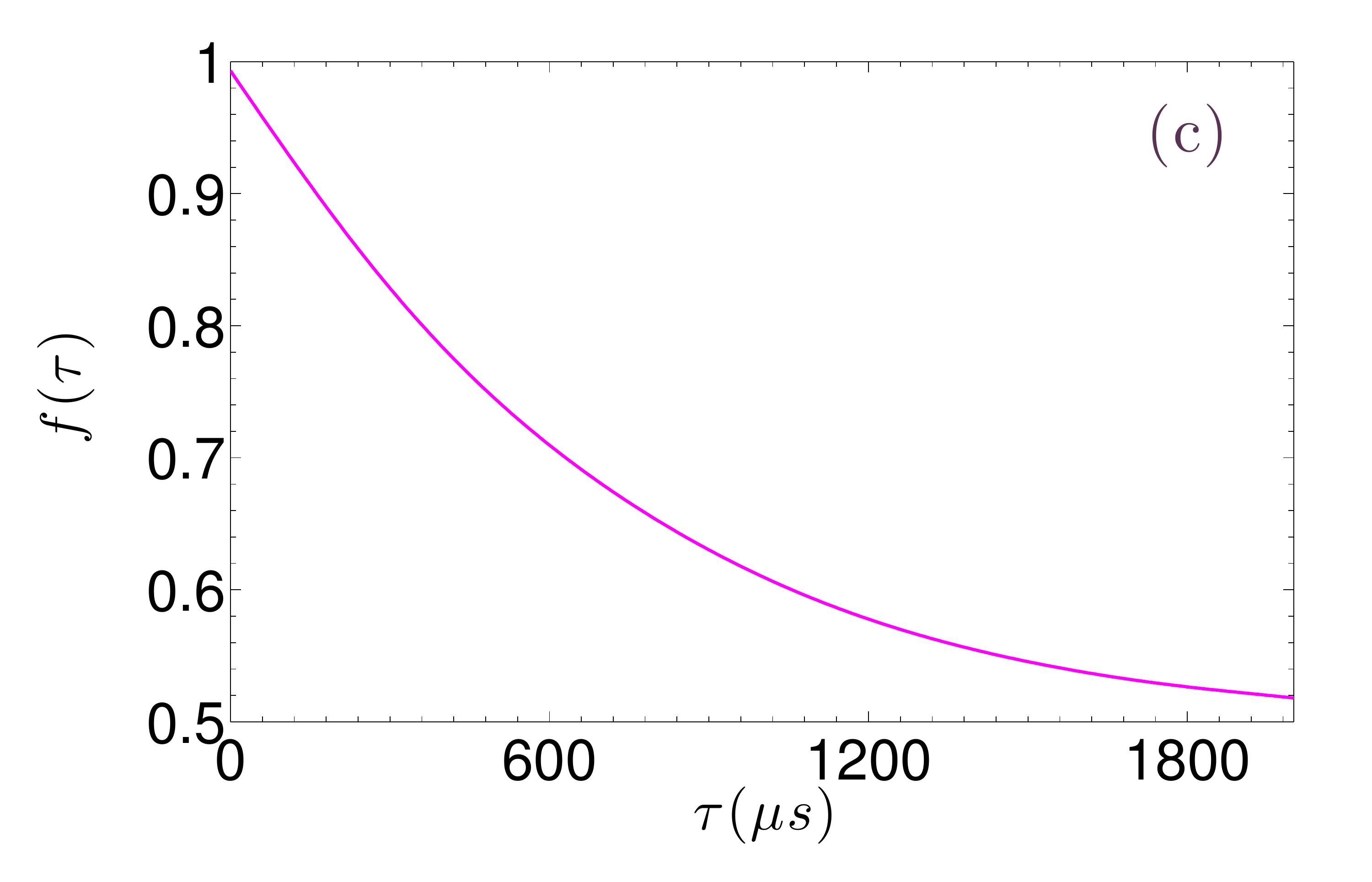}
\hspace{-0.2cm}
\includegraphics[width=4.5cm]{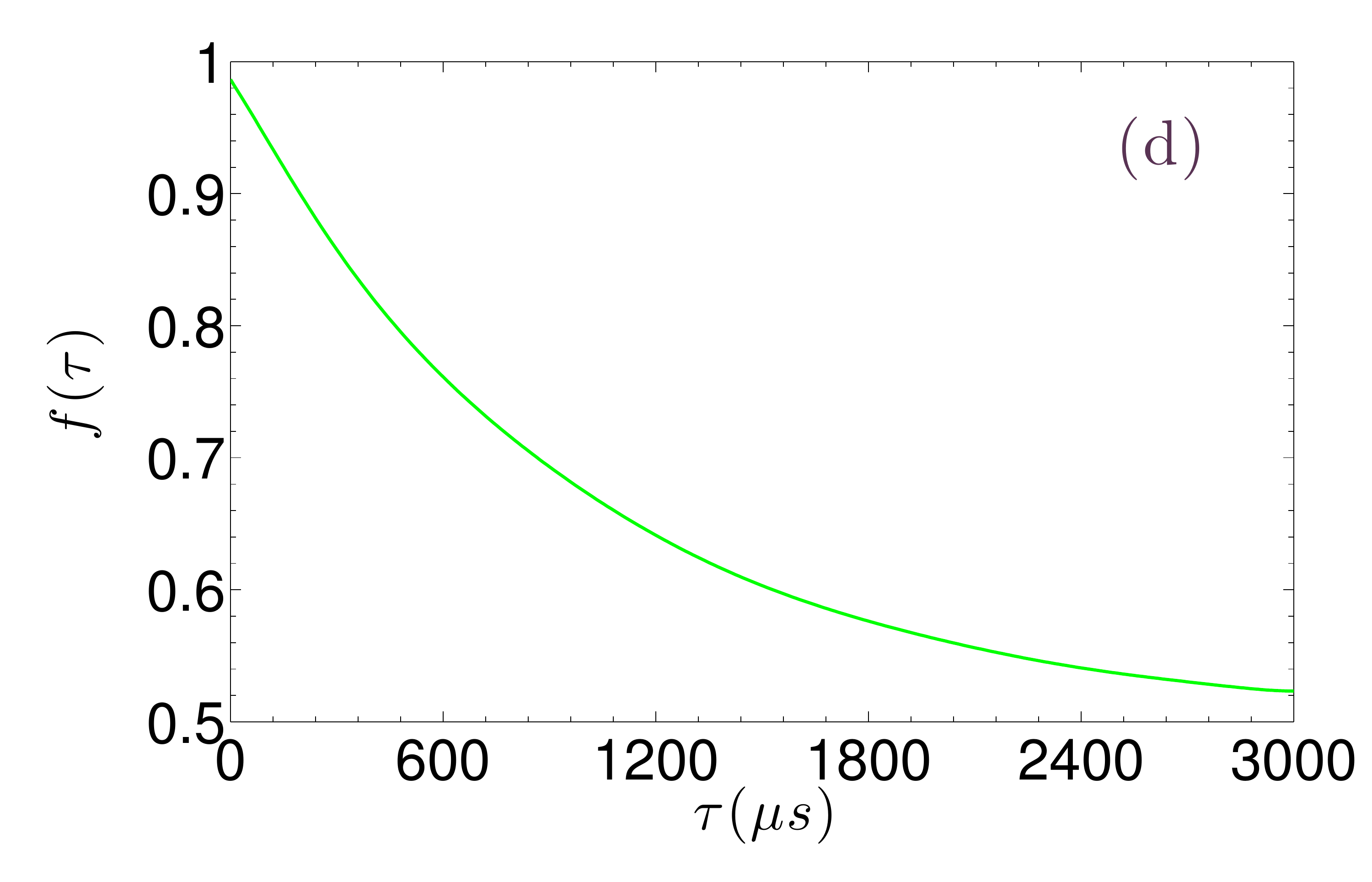}
\end{minipage}
\end{center}
\caption{(Color online) Extrapolation curves from numerical simulations for the decay of coherence $f(\tau)$ with the first- (a), second-(b), third-(c) and fourth-order (d) continuous dynamical decoupling.The Rabi frequency is $\Omega_1=40 \mbox{MHz}$ and $\Omega_{K}=\Omega_{K-1}/30$ for $K=2,3,4$, and the parameter $\Gamma$ is chosen for $\mbox{T}_1=1.5 \mbox{ms}$. }\label{HIOD-s}
\end{figure}

We model the magnetic noise and the microwave fluctuation by Ornstein-Uhlenbeck processes \cite{Hanson08}. The system dynamics is described by the following quantum master equation
\begin{equation}
\frac{d}{dt}\rho =-i[H(t),\rho]+ \frac{\Gamma}{2}\sum_{a=\sigma_\pm}(2a^{\dagger} \rho a -\rho a a^{\dagger} -a a^{\dagger}\rho)
\label{ME-s}
\end{equation}
where $H(t)$ is the total Hamiltonian including both the magnetic noise and the driving fields with fluctuations. The Lindblad operators represent the relaxation process with the rate $\Gamma$ corresponding to $\mbox{T}_1=1.5 \mbox{ms}$ which is close to the value of the diamond used in our experiment. The relaxation in the above master equation takes the high temperature, which is valid for our experiment at room temperature. We generate $2000$ realizations of Ornstein-Uhlenbeck processes with the exact simulation algorithm \cite{Gill96} for the magnetic noise $\delta_b(t)$ and the microwave fluctuations $\delta_i(t)$. We choose the correlation time for the spin bath as $\tau_c=25\mu s$. The fluctuation of microwave filed amplitude is usually much slower and we use the value of correlation time $\tau_m=1 \mbox{ms}$. We first simulated numerically the experimental data in Fig.\ref{HIOD-s}(a) for the case of Rabi oscillation with one single driving field. The numerical result agrees well with the experimental data and the Gaussian fit of the decay envelope (arising from slow noise \cite{KuboBook}) $S_1(\tau)=\exp{(-b_1^2\tau^2/2)}$ (see Fig.2(a) of the main text). This supports that an Ornstein-Uhlenbeck process serves as a good model for the driving field fluctuation. We estimate that the relative amplitude fluctuation of the driving field is about $2.4\times 10^{-3}$. We use these estimated parameters for the numerical simulations of higher-order driving schemes by solving the master equation in Eq.(\ref{ME-s}). In Fig.\ref{HIOD-s}, we plot the coherence decay of the dressed qubit quantified by
\begin{equation}
f(\tau)=\left \vert \left\langle \psi_I(0) \vert \psi_I(\tau) \right \rangle\right \vert
\label{ftau-s}
\end{equation}
with $\ket{\psi_I(0) } =\ket{\downarrow}$ which is a coherent superposition of the dressed qubit ($\ket{\uparrow}_{x/y}$ and $\ket{\downarrow}_{x/y}$), and $\ket{\psi_I(\tau)}$ is the state of the dressed qubit after time $\tau$. The results in Fig.\ref{HIOD-s} (see also Fig.3(d) in the main text) shows the scaling of the coherence time with high-order driving and the possibility to approach $\mbox{T}_1$. We remark that for the first- and second-order drivings, the decay envelop is more like Gaussian, while for third- and fourth-order drivings the decay envelop becomes exponential (arising from the relaxation) as the effect of microwave fluctuations is increasingly suppressed.
%

%\newpage


\begin{thebibliography}{99}
\bibitem{footnote} In our experiment, we use the simplified second-order field as $H_{d_2}=\hbar\Omega_2  \cos{\blb{\bla{\omega+\Omega_1}t+\varphi}}\sigma_x$, which leads to a similar second-order effective Hamiltonian. The estimated improvement is however slightly smaller than those obtained for the numerical simulations using the driving fields in Eq.(\ref{Hd2-m}) due to the presence of counter-rotating terms that are cancelled for the fields in Eq.(\ref{Hd2-m}).
\end{thebibliography}
\end{document}